\documentclass[twocolumn]{aastex631}
\usepackage{amsmath}
\usepackage{enumitem}
\usepackage{xspace}

\undef\listofchanges
\usepackage[commandnameprefix=always,commentmarkup=uwave]{changes}
\definechangesauthor[color=cyan]{wf}

\newcommand{\kms}{\ensuremath{\,\rm{km}\,\rm{s}^{-1}}\xspace}

\newcommand{\Rsun}{\ensuremath{\,\rm{R}_{\odot}}}
\newcommand{\Msun}{\ensuremath{\rm{M}_{\odot}}}

\newcommand{\Zsun}{\ensuremath{\,\rm{Z}_{\odot}}}

\newcommand{\Gyr}{\ensuremath{\,\rm{Gyr}}}

\newcommand{\fe}{\ensuremath{\eta_{\rm{form}}}\xspace}
\newcommand{\feNSNS}{\ensuremath{\eta_{\rm{form, NSNS}}}\xspace}
\newcommand{\feBHNS}{\ensuremath{\eta_{\rm{form, BHNS}}}\xspace}
\newcommand{\feBHBH}{\ensuremath{\eta_{\rm{form, BHBH}}}\xspace}

\newcommand{\COMPAS}{{\tt COMPAS}\xspace}

\newcommand{\compasv}{\texttt{v03.01.02}\xspace}

\definecolor{brightmaroon}{rgb}{0.7, 0.09, 0.09} 

\newcommand{\edited}[1]{#1}

\definecolor{citeblue}{rgb}{0.26, 0.53, 0.96} 
\newcommand{\tocite}[1]{{\color{black}#1}}

\received{\today}

\begin{document}

\title{Not just winds: why models find binary black hole formation is metallicity dependent, while binary neutron star formation is not}
\shorttitle{Why BHBH formation is Z dependent while NSNS formation is not}

\correspondingauthor{L.~van Son}
\email{lvanson@flatironinstitute.org}

\author[0000-0001-5484-4987]{L.~A.~C.~van~Son}
\affiliation{Center for Computational Astrophysics, Flatiron Institute, 162 Fifth Avenue, New York, NY 10010, USA}
\affiliation{Department of Astrophysical Sciences, Princeton University, 4 Ivy Lane, Princeton, NJ 08544, USA}

\author[0000-0001-9295-5119]{S. K. Roy}
\affiliation{Department of Physics and Astronomy, Stony Brook University, Stony Brook NY 11794, USA}

\author[0000-0002-6134-8946]{I. Mandel}
\affiliation{School of Physics and Astronomy, Monash University, Clayton VIC 3800, Australia}
\affiliation{ARC Center of Excellence for Gravitational-wave Discovery (OzGrav), Melbourne, Australia}

\author[0000-0003-1540-8562]{W. M. Farr}
\affiliation{Department of Physics and Astronomy, Stony Brook University, Stony Brook NY 11794, USA}
\affiliation{Center for Computational Astrophysics, Flatiron Institute, 162 Fifth Avenue, New York, NY 10010, USA}

\author[0009-0003-9165-9889]{A. Lam}
\affiliation{Graduate Center, City University of New York, 365 5th Avenue, New York, NY 10016, USA}

\author[0000-0000-0000-0000]{J. Merritt}
\affiliation{Institute for Fundamental Science, Department of Physics, University of Oregon, Eugene, OR 97403, USA}

\author[0000-0002-4421-4962]{F. S.~Broekgaarden}
\affiliation{Department of Astronomy \& Astrophysics, University of California, San Diego, 9500 Gilman Drive, La Jolla, CA 92093, USA}

\author[0000-0002-2090-9751]{A.A.C. Sander}
\affiliation{Zentrum f{\"u}r Astronomie der Universit{\"a}t Heidelberg, Astronomisches Rechen-Institut, M{\"o}nchhofstr. 12-14, 69120 Heidelberg, Germany}

\author[0000-0001-5261-3923]{J. J. Andrews}
\affiliation{Department of Physics, University of Florida, 2001 Museum Rd, Gainesville, FL 32611, USA}
\affiliation{Institute for Fundamental Theory, 2001 Museum Rd, Gainesville, FL 32611, USA}

\begin{abstract}
Both detailed and rapid population studies alike predict that binary black hole (BHBH) formation is orders of magnitude more efficient at low metallicity than high metallicity, while binary neutron star (NSNS) formation remains mostly flat with metallicity, and black hole-neutron star (BHNS) mergers show intermediate behavior.
This finding is a key input to employ double compact objects as tracers of low-metallicity star formation, as spectral sirens, and for merger rate calculations. Yet, the literature offers various (sometimes contradicting) explanations for these trends.
We investigate the dominant cause for the metallicity dependence of double compact object formation. 
We find that the BHBH formation efficiency at low metallicity is set by initial condition distributions, and conventional simulations suggest that about \textit{one in eight interacting binary systems} with sufficient mass to form black holes will lead to a merging BHBH. 
We further find that the significance of metallicities in double compact object formation is a question of formation channel.
The stable mass transfer and chemically homogeneous evolution channels mainly diminish at high metallicities due to changes in stellar radii, while the common envelope channel is primarily impacted by the combined effects of stellar winds and mass-scaled natal kicks. 
Outdated giant wind prescriptions exacerbate the latter effect, suggesting BHBH formation may be much less metallicity dependent than previously assumed. 
NSNS formation efficiency remains metallicity independent as they form exclusively through the common envelope channel, with natal kicks that \edited{are assumed} uncorrelated with mass.
Forthcoming GW observations will provide valuable constraints on these findings.
\end{abstract}


\section{Introduction} \label{sec:intro}
We are nearing the conclusion of the first decade of gravitational-wave (GW) astronomy \citep{GWTC1_pop2019ApJ...882L..24A,GWTC2_pop2021ApJ...913L...7A,GWTC3_pop2023PhRvX..13a1048A}. 
The rapidly growing population of GW sources provides new opportunities to understand the lives and deaths of otherwise elusive massive stars.
However, making any statements about the nature of their progenitor stars is very challenging due to the many uncertainties associated with the evolution of massive stars \citep[e.g.][]{2024ARA&A..62...21M}. 
These numerous modeling uncertainties lead to large variations in the predicted properties and merger rates of double compact object mergers \citep[e.g.][]{2022LRR....25....1M}.
It is thus rather remarkable when (rapid) population synthesis models converge on certain results, especially when this is achieved under various assumptions and input physics.

\begin{figure*}
    \centering
    \includegraphics[width=\textwidth]{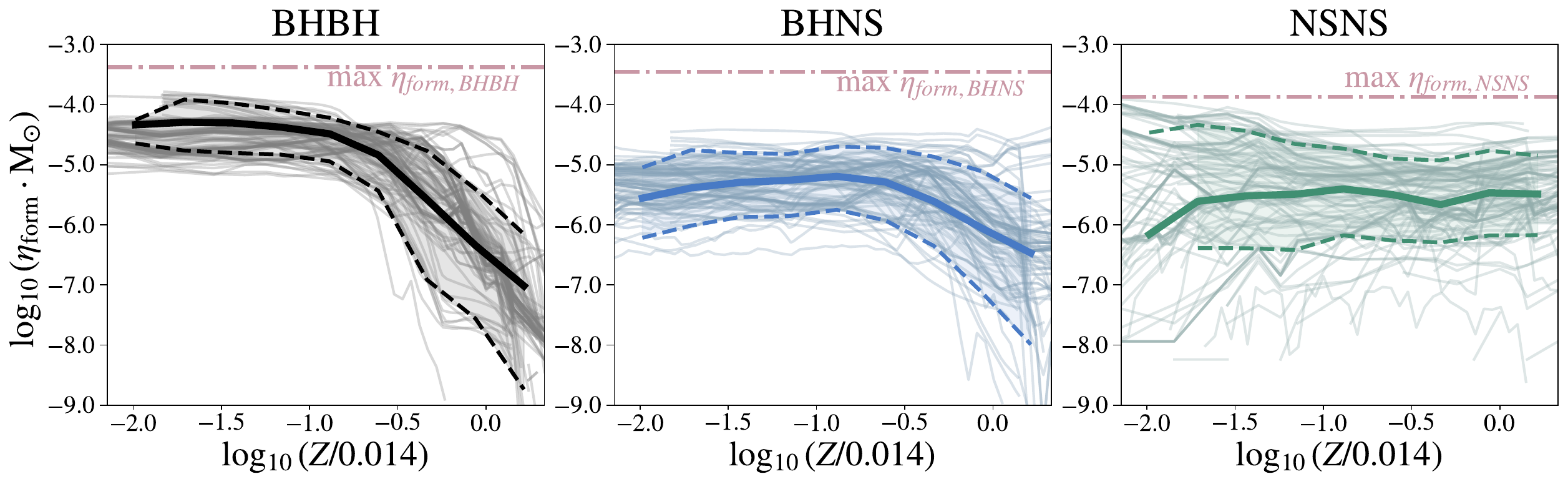}
    \caption{The formation efficiency of BHBH, BHNS, and NSNS as a function of the metallicity from different studies 
    \citep{2018MNRAS.480.2011G,2018A&A...619A..77K,2019MNRAS.490.3740N,2021MNRAS.502.4877S,2022MNRAS.516.5737B,2023MNRAS.524..426I}.
    The solid and dashed lines show the median, 10th and 90th percentile, respectively. 
    The BHBH yield is consistently flat at low metallicity, and declines by multiple orders of magnitude for $Z \gtrsim 0.2\Zsun$. BHNS formation exhibits a similar trend, albeit to a lesser extent, while NSNS formation shows minimal to no dependence on metallicity.
    Pink dash-dotted lines show the theoretical maxima from equations \ref{eq: max eta bhbh},  \ref{eq: max eta nsns}, and \ref{eq: max eta bhns}, which we find to be surprisingly close to some of the simulation results. }
    \label{fig: yield review}
\end{figure*}

Over the past decade, numerous studies have consistently found similar results regarding the `yield' or `formation efficiency' of Binary Black Hole (BHBH), Black Hole-Neutron Star (BHNS), and Neutron Star-Neutron Star (NSNS) mergers as a function of metallicity (see Figure \ref{fig: yield review} for a review). 
We define formation efficiency as the number of double compact object (DCO) systems formed per unit of star formation, that have a delay time \added[id=wf]{to merger} shorter than the present-day age of the Universe.  
In other words, the number of DCO `mergers' formed per solar mass of star formation at a certain metallicity is
\begin{equation}
    \fe(Z) =  \frac{\mathrm{d}N_\mathrm{DCO}(Z)}{\mathrm{d} M_\mathrm{SF}},
\end{equation}
with units of per \Msun. 
The effect of metallicity on single stars is two-fold. 
Firstly, radiatively-driven winds (like main sequence and Wolf-Rayet winds) are stronger at higher metallicity \citep[e.g.,][]{2001A&A...369..574V,2005A&A...442..587V}, thereby lowering the star's mass and final core/remnant formation mass \citep[see discussions in e.g., ][]{2010ApJ...714.1217B,2017A&A...603A.118R, 2020ApJ...896...56W, 2024arXiv240707204V, 2023NatAs...7.1090B}. %
Second, stellar radii increase with metallicity due to higher opacity which makes stars more ‘puffed up’. However, the maximum radius at different metallicities is uncertain and depends strongly on model assumptions  \citep[see, e.g.,][]{2023MNRAS.525..706R}.
\edited{Moreover, at low metallicities, incomplete stripping during mass transfer can cause the stripped star to expand again \citep[potentially leading to additional mass transfer phases][]{2020A&A...637A...6L}, introducing additional metallicity dependence of (stripped-star) radii. However, this effect is not captured by typical rapid population synthesis codes.}
Both the wind and radius effects can significantly impact the formation efficiency of double compact object mergers.

The impact of metallicity on binary evolution and ultimately the formation efficiency of double compact object mergers has been extensively studied using a wide range of binary physics parameters and codes.
Specifically, using BSE~\citep{2002MNRAS.329..897H} based codes including \texttt{MOBSE} (fig. 12 from~\citealt{2018MNRAS.474.2959G}, fig. 14 from~\citealt{2018MNRAS.480.2011G}, fig. 1 from~\citealt{2021MNRAS.502.4877S}),  \texttt{StarTrack} (fig. 6 from \citealt{2018A&A...619A..77K}, and \citealt{2010ApJ...715L.138B, 2013ApJ...779...72D, 2016Natur.534..512B}),  \texttt{COMPAS} (fig. 1 from \citealt{2019MNRAS.490.3740N}, fig. 1 from \citealt{2022MNRAS.516.5737B}, and fig. 4 from \citealt{2023ApJ...957L..31F} using models from \citealt{2022ApJ...940..184V}).
Furthermore, this has been studied using \texttt{BPASS} \citep[e.g., Fig. 4 from ][]{2016MNRAS.462.3302E}, which employs a version of the Cambridge STARS stellar evolution code to compute stellar evolution on-the-fly, and \texttt{SEVN} (figures 17 and 18 from \citealt{2023MNRAS.524..426I}), which interpolates main stellar-evolution properties from pre-computed tables based on PARSEC. 
All models displayed in Figure \ref{fig: yield review} agree on several key trends:

\paragraph{The BHBH formation efficiency} plateaus at low metallicity ($Z$) and drops drastically at higher metallicity, typically above $Z \approx 0.2\Zsun \approx 0.003$.
Several explanations have been proposed to account for the decline in BHBH formation efficiency at high metallicity:
\begin{enumerate}[nolistsep]
    \item At high $Z$, more stellar winds:\\
    A) result in less massive carbon-oxygen cores, causing larger natal kicks that unbind the binary system
    \citep{2018A&A...619A..77K,2019MNRAS.490.3740N,2022MNRAS.516.5737B}.

    B) result in less massive carbon-oxygen cores, thus requiring higher-mass ZAMS stars to create the same mass BH
    \citep{2010ApJ...715L.138B,2010ApJ...714.1217B,2016Natur.534..512B,2024AnP...53600170C, 2016MNRAS.462.3302E}.
    In other words, there is a metallicity-dependent maximum BH mass that increases towards low $Z$ \citep[e.g.,][]{2021MNRAS.504..146V,2024ApJ...964L..23R,2024MNRAS.529.2980W}.

    C) cause wider (and lower-mass) systems at double compact object (DCO) formation, causing longer coalescence times \citep{Peters:1964}, and thus prevent mergers
    \citep{2016MNRAS.462.3302E,2023MNRAS.524..426I}.

    D) \edited{remove so much mass that they} keep stars more compact in size, thereby avoiding interaction 
    \citep{2023MNRAS.524..426I}.

    \item At high $Z$, stars will have a larger radius while on the Hertzsprung gap (HG). 
    A common envelope instigated from a donor that is on the HG is typically assumed to lead to a stellar merger \citep[e.g.,][]{2012ApJ...759...52D}. 
    Assuming that BHBH formation is dominated by the common envelope (CE) channel, this means that the formation efficiency would drop at high $Z$ 
    \citep{2010ApJ...715L.138B, 2017MNRAS.472.2422M, 2018MNRAS.480.2011G, 2018A&A...619A..77K,2022MNRAS.516.5737B}.
    
\end{enumerate}

\paragraph{The BHNS formation efficiency} plateaus at low $Z$, similar to BHBH, though generally at a lower absolute value. It also decreases at the highest $Z$, though less drastically than BHBH. In several variations, it displays a shallow peak around $0.2\Zsun$.
It is commonly assumed that the same factors affecting BHBH formation efficiency also influence BHNS formation efficiency \citep{2018A&A...619A..77K,2022MNRAS.516.5737B,2023MNRAS.524..426I}.
The plateau in formation efficiency at low metallicity is argued to be lower for BHNS than for BHBH for the following reasons:
\begin{enumerate}[nolistsep]
    \setcounter{enumi}{2}
    \item Only low-mass BHs are found in BHNS. At low metallicities, there is thus less to be gained from the initial mass function, causing the \textit{lower} plateau value \citep{2022MNRAS.516.5737B}. This is in essence the analog to argument 1B. 
    
    \item The BHNS systems receive at least one natal kick (at the formation of the NS), causing many more systems to be disrupted \citep{2018A&A...619A..77K}.
    
    \item The formation channel where two common envelopes are required becomes inefficient at low $Z$ and higher mass stars, because for these kinds of stars, common envelopes are less common and more often lead to stellar mergers (\citealt{2018A&A...619A..77K}, see also \citealt{2022ApJ...931...17V}).
\end{enumerate}

Notably, \cite{2023MNRAS.524..426I} observe an \textit{increase} in BHNS formation efficiency with metallicity in models that vary the envelope binding energy ($\lambda_{\mathrm{CE}}$, their models LX, LC, and LK).

\vspace{-0.3cm}

\paragraph{The NSNS formation efficiency} remains mostly independent of Z but is highly sensitive to the chosen physics assumptions.
One explanation for this behavior is that the progenitors of NSNS systems are typically lower-mass stars, which undergo less mass loss overall \citep[e.g.,][]{2019MNRAS.490.3740N}. 
Note that the NSNS formation efficiency is not necessarily constant with metallicity \citep[see e.g.][]{2023ApJ...955..133G}, but rather does not show the consistent drop at high metallicity observed in BHBH and BHNS formation. 
Instead, NSNS formation appears to be more influenced by binary physics assumptions than by metallicity. 
Variations in the common envelope parameters can cause changes in the NSNS formation efficiency by several orders of magnitude \citep[e.g.,][]{2017MNRAS.472.2422M,2018MNRAS.480.2011G,2021MNRAS.502.4877S,2022MNRAS.516.5737B}, which explains the large spread in Figure \ref{fig: yield review}. \\

These arguments will be referenced throughout this work as we systematically investigate the dominant physical processes driving the metallicity dependence of DCO formation efficiency.

Given that these behaviors are some of the most consistent results across population synthesis codes, they warrant closer inspection and scrutiny. Additionally, the robustness of these findings carries several significant implications. 
Firstly, the majority of cosmic star formation occurs at relatively high metallicities.
Although the exact shape of the metallicity-dependent cosmic star formation is uncertain \citep[e.g.,][]{2024AnP...53600170C}, most models agree that less than $30\%$ of stellar mass is formed at low metallicity ($Z < \Zsun/10$) in the past $\sim11.5\Gyr$ (redshifts below 3, see \citealt{2021MNRAS.508.4994C}, and figure 6 in \citealt{2023ApJ...948..105V}). 
Consequently, even a slight shift in the metallicity at which the drop in $\feBHBH$ occurs can lead to substantial (potentially many orders of magnitude) changes in the predicted merger rate \citep[e.g.,][]{2019MNRAS.490.3740N,2022MNRAS.516.5737B,2021MNRAS.502.4877S,2024AnP...53600170C}.
Understanding the metallicity dependence of formation efficiencies is thus crucial to reduce the uncertainty in the predicted DCO merger rate.
The metallicity dependence of BHBH formation is therefore also intimately intertwined to claims like `the (heavy) BHBH mergers we observe today must come from the early, low-metallicity Universe' \citep{2016PhRvL.116f1102A,2016Natur.534..512B,2017NatCo...814906S,2023ApJ...957L..31F}.

Second, because the BHBH formation efficiency depends so strongly on metallicity, it has been proposed as a tracer of low-metallicity star formation, which is otherwise challenging to constrain \citep[see][and references therin]{2024AnP...53600170C}. 
Hence, it is essential to understand if and why this low-metallicity formation efficiency prediction is robust.

Finally, the relative independence of NSNS formation efficiency on metallicity (combined with typically shorter delay times and a smaller range of final masses) suggests that the NSNS mass distribution will vary much less with redshift with respect to the BHBH mass distribution. 
This implies that NSNS systems could serve as `spectral sirens' \citep{Chernoff_1993, Taylor_2012, Taylor_2012_1, 2019ApJ...883L..42F,2022PhRvL.129f1102E}, and help infer cosmological parameters \citep{2024arXiv241102494R}.  \\

In this paper, we set out to answer the following questions: 
I) What determines the low-metallicity plateau height in the BHBH and BHNS formation efficiencies?
II) What is the dominant physics responsible for the metallicity-dependent behavior of formation efficiency?
III) How robust is this behavior?
Section \ref{sec: max form eff} addresses question I).  
In Sections \ref{sec: methods} and \ref{sec: dominant reason to fail} we describe our method and systematically evaluate the different proposed explanations for the metallicity (in)dependence of double compact object formation, thereby addressing question II).
We discuss the robustness of these relations (question III) in Section \ref{sec: phys var}, and summarize in Section \ref{sec: conclusion}.

\section{Theoretical maximum formation efficiency}\label{sec: max form eff}
The (nonsensical) upper limit of $\fe = 1/\Msun$, would mean that there is one merging DCO for every unit of stellar mass formed.  This is of course impossible as more than $1\Msun$ is needed to create a single NS or BH, let alone a DCO.  
A more useful maximum considers the formation efficiency when all stars that \textit{can} potentially form merging double compact object binaries, would indeed do so. 
The goal of this exercise is to examine the plateau observed in simulations at low metallicities for both BHBH and BHNS (see Figure \ref{fig: yield review}), and evaluate how far this lies from the maximum yield if we ignore the complexities of binary interactions \citep[i.e., require only that stars are not effectively single cf.][]{2012Sci...337..444S,2023ASPC..534..275O}. 
To achieve this, we use a Drake-like equation for formation efficiency, similar to \cite{2022PhR...955....1M}, and \cite{2022LRR....25....1M}: 
\begin{equation}
\begin{split}
     \fe = \frac{1}{\langle M_{SF} \rangle } \Bigl( f_{\mathrm{primary}} \times f_{\mathrm{secondary}} \times f_{\mathrm{init \, sep}} \\
     \times f_{\mathrm{survive \, SN1}} \times f_{\mathrm{survive \, SN2}} \Bigr)
\end{split}
\label{eq: drake} 
\end{equation}

Here, $f_{\mathrm{primary}}$ represents the fraction of the initial primary masses that are high enough to form a compact object, and $f_{\mathrm{secondary}}$ is the fraction of that population with a companion mass sufficiently high to also form the compact object of interest. $f_{\mathrm{init \, sep}}$ represents the fraction of the population born at the `right' separation, while $f_{\mathrm{survive \, SN1}}$ and $f_{\mathrm{survive \, SN2}}$ are the fractions of systems that survive the first and second supernova (SN) natal kicks, respectively.
We divide this by the average star forming mass per system formed, $\langle M_{SF} \rangle$, to transform this into a formation efficiency. 
By assuming a Kroupa IMF \citep{2001MNRAS.322..231K} that ranges between $0.01\Msun - 300\Msun$ \citep{2013A&A...558A.134D}, and a mass-dependent binary fraction following \cite{2023ASPC..534..275O} we get $\langle M_{SF} \rangle \approx 0.51\Msun$ (see Appendix \ref{app: drake} for more details).

\begin{figure}
    \centering
        \includegraphics[width=0.48\textwidth]{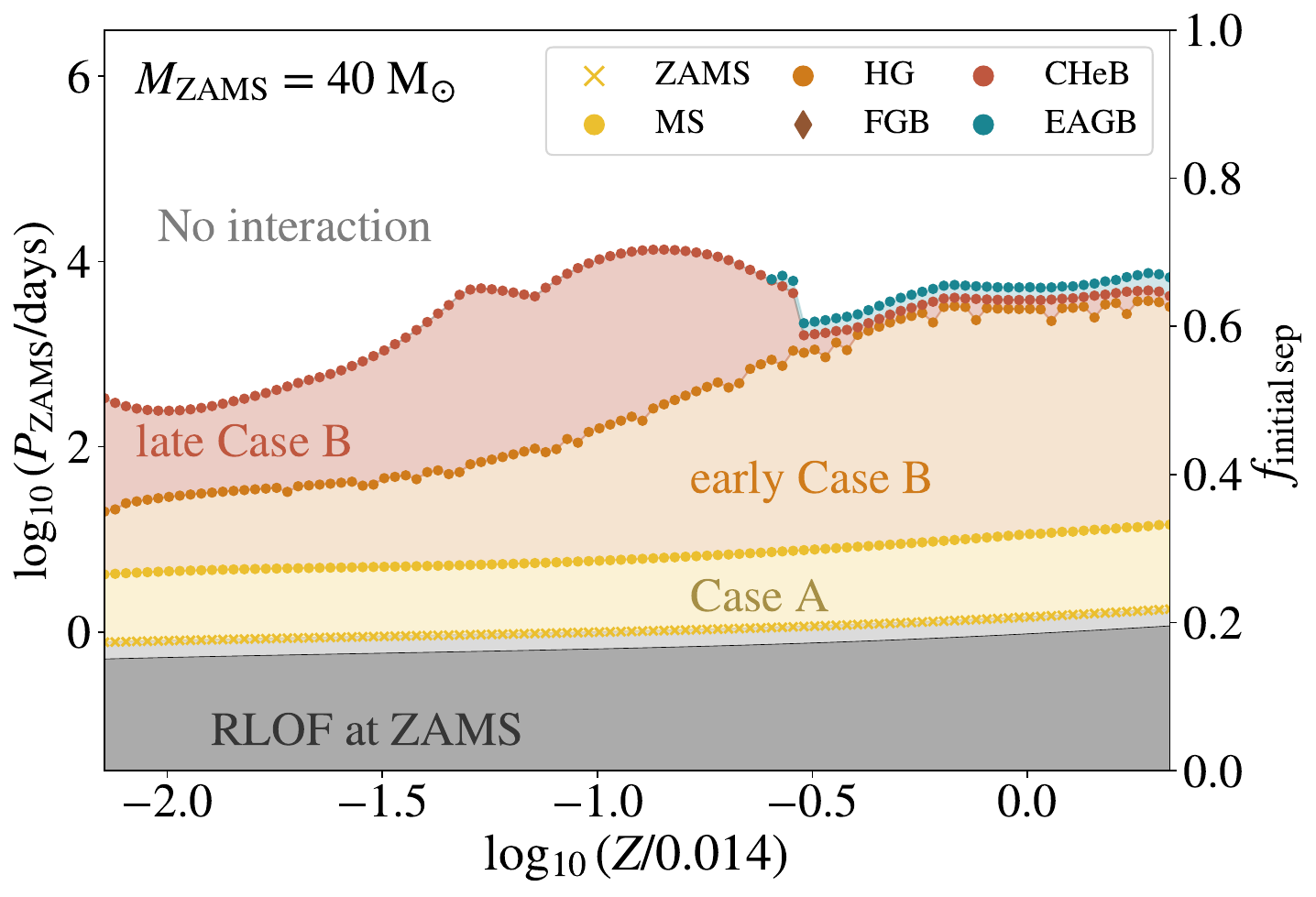}
        \includegraphics[width=0.48\textwidth]{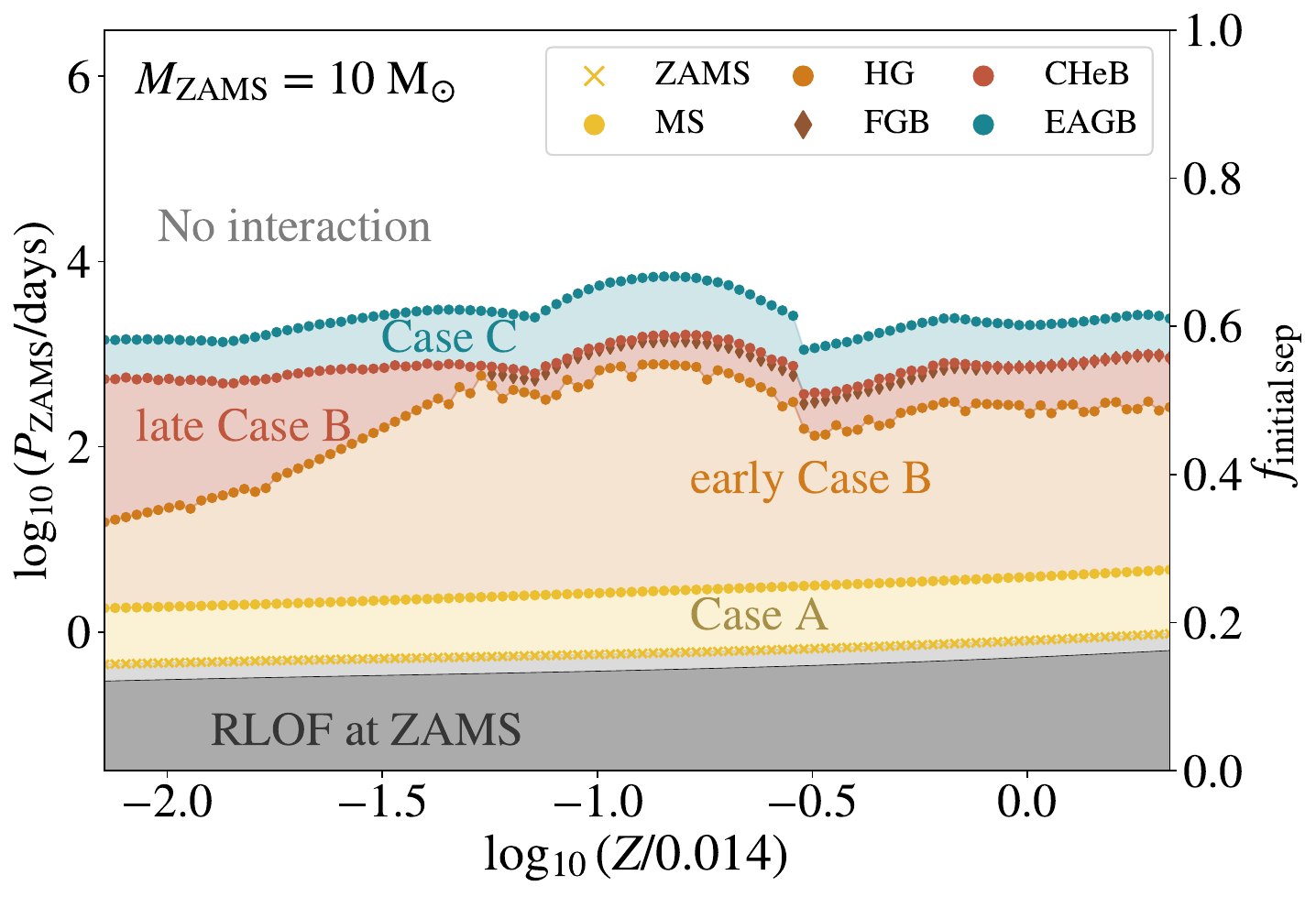}
    \caption{\edited{The first mass transfer that a star will encounter as a function of metallicity, given its ZAMS orbital period and assuming an equal mass companion.}
    Scatter points indicate the stellar evolutionary type for a typical BH progenitor ($M_{\mathrm{ZAMS}} = 40\Msun$, top), and a typical NS progenitor ($M_{\mathrm{ZAMS}} = 10\Msun$, bottom).
    Shading indicates whether mass transfer would be classified as Case A, B or C. 
    Gray shading indicates RLOF at ZAMS; dark gray shows where this would lead to a stellar merger. 
    See text for acronyms.
    We find $f_{\mathrm{init \, sep}} \approx 0.4$ for NSs and BHs. 
    \label{fig: SSE properties} }
\end{figure}

To estimate $f_{\mathrm{primary}} \times f_{\mathrm{secondary}}$, we compute the fraction of the initial mass function (IMF) that forms compact objects. Using a Kroupa IMF \citep{2001MNRAS.322..231K} and a minimum ZAMS mass of $20\Msun$ for BH formation gives $f_{\mathrm{primary}} \approx 4 \times 10^{-4}$. 
This value remains nearly the same when using the COMPAS rapid population synthesis code \citep[version \compasv][]{2017NatCo...814906S,2018MNRAS.481.4009V,2022ApJS..258...34R}, as shown in Figure \ref{fig: rem masses}, in Appendix \ref{app: single stars} where $f_{\mathrm{primary}} \approx 1 \times 10^{-3}$ for BHs. 
Note that this contradicts 1B from Section \ref{sec:intro}, since the change in $f_{\mathrm{primary}}$ with metallicity changes more for NS formation than for BH formation, suggesting that \feNSNS should also increase for lower $Z$. 
To properly calculate $f_{\mathrm{primary}} \times f_{\mathrm{secondary}}$, we consider the \textit{joint} probability of selecting both masses in the $20-300\Msun$ range, assuming a uniform mass ratio distribution \citep{Sana:2012Sci, 2017ApJS..230...15M}. 
This results in $f_{\mathrm{primary}} \times f_{\mathrm{secondary}} \approx 5.05 \times 10^{-4}$ (see Appendix \ref{app: drake}). 

Next, we would like to know what fraction of the binaries leads to the right interactions, such that they produce the compact systems that allow for GW emission. In reality, this is a complex function that encompasses all of binary interactions, and depends on the adopted physics. 
However, to zeroth order, we can require that binaries must interact in order to shrink their orbits \citep[though cf. formation with `lucky SN kicks' in e.g.,][]{2002ApJ...572..407B,2021MNRAS.508.5028B}.
To compute the maximum possible yield, we set $f_{\mathrm{init \, sep}}$ to be the fraction \edited{\textit{interacting}} binaries that do not merge at ZAMS.
Figure \ref{fig: SSE properties} shows the period of first mass transfer for a typical NS ($M_{\mathrm{ZAMS}} = 10\Msun$, top), and BH progenitor ($M_{\mathrm{ZAMS}} = 40\Msun$, bottom) assuming an equal mass companion. 
\footnote{For completeness, we show the maximum radii per $Z$ for a range of stellar masses in Appendix \ref{app: single stars} }
The scatter points show the stellar type at the onset of mass transfer: Zero Age Main Sequence (ZAMS),  Main Sequence (MS), Hertzsprung Gap (HG), First Giant Branch (FGB), Core Helium Burning (CHeB), or Early Asymptotic Giant Branch (EAGB). 
We assume that stars that experience RLOF at ZAMS lead to stellar mergers.  
We adopt a flat-in-log distribution of initial separations between $0.01$ and $1000\,\mathrm{AU}$. 
Using Figures \ref{fig: SSE properties}, and \ref{fig:max R per Z} we estimate separations avoiding mergers to be roughly $\approx 0.1 \mathrm{AU}$ and the maximum for interaction to be approximately $14\mathrm{AU}$.
This leads to $f_{\mathrm{init , sep}} \approx 0.43$.

Lastly, both stars might receive a natal kick at compact object formation. 
The kick distribution of black holes remains a topic of debate \citep[see e.g, ][for a review]{2016arXiv160908411M}, however, at least some black holes appear to be born without a natal kick (see Section \ref{ss: obs BH kicks}).
The most optimistic assumption is for BHs to receive no kick, setting $f_{\mathrm{survive \, SN1}} = f_{\mathrm{survive \, SN2}} = 1$. 

Combining this, we find that the expected maximum formation efficiency for BHBHs is about 
\begin{equation}\label{eq: max eta bhbh}
\begin{split}
    \feBHBH & = \frac{5.06 \cdot 10^{-4}}{0.52\Msun}  \times 0.43 \times 1  \times 1 \\ 
& \approx 4.21 \cdot 10^{-4} \Msun^{-1}.
\end{split}
\end{equation}

Remarkably, this is only about eight times higher than the median \feBHBH, and just four times higher than the $90^{\rm th}$ percentile at low metallicities, as shown in Figure \ref{fig: yield review}.\footnote{Note that the median in Figure \ref{fig: yield review} is not a `true' median, since different draws from the same study/work are not completely independent. }
This implies that the studies reviewed in Figure \ref{fig: yield review} suggest about \textit{one in every eight} binary systems, with masses high enough to form a black hole and separations small enough to interact, will lead to a merging binary black hole. 
\edited{This is surprising; we would expect simulations to predict values far below equation \ref{eq: max eta bhbh} since the latter assumes \textit{none} of the binary evolution leads to stellar mergers nor unbound systems, which is unrealistic. }

We repeat this exercise to estimate the maximum formation efficiency for NSNS, now assuming $8\Msun < M_{\mathrm{ZAMS}} \leq 20\Msun$ to form a NS. 
We find $f_{\mathrm{primary}} \times f_{\mathrm{secondary}} \approx 8.18 \times 10^{-4}$.
Neutron stars are generally assumed to receive natal kicks following a Maxwellian distribution with $\sigma_{\mathrm{kick}}=65\kms$ \citep[motivated by pulsar proper motion observations][]{2005MNRAS.360..974H}. 
Electron-capture SN may also form NSs, typically with smaller natal kicks \citep[$\sigma_{\mathrm{kick}} = 30 \kms$,][]{2015MNRAS.454.3073S, 2018ApJ...865...61G, 2019MNRAS.484.3307M}. 
Population studies estimate post-SN binary survival rates, with \cite{2019A&A...624A..66R} finding $f_{\mathrm{survive , SN1}} = 0.14^{+.22}_{-.10}$ and \tocite{Lam et al.\ (in prep)} predicting $f{\mathrm{survive , SN1}}$ between $0.09$ and $0.23$. 
The variation in these estimates arises from differing assumptions on e.g., the mass transfer physics and kick distributions. 
In both studies, the upper limits stem from assuming $\sigma_{\mathrm{kick}}\approx30\kms$, for part or all of compact object formation. 
We choose to adopt the upper limit from \tocite{Lam et al.\ (in prep)} for $f_{\mathrm{survive, SN1}} \approx 0.23$.
The value for $f_{\mathrm{survive \, SN2}}$ is even more complex, as the binary has typically experienced two phases of mass transfer, and its orbital parameters at this stage are highly sensitive to the intricacies of binary evolution. 
However, generally simulations predict that $f_{\mathrm{survive \, SN2}}$ is much higher than $f_{\mathrm{survive \, SN1}}$ \citep[e.g.][]{2018MNRAS.481.4009V}. 
We thus adopt $f_{\mathrm{survive \, SN2}} = 1.0$, and find that 

\begin{equation}\label{eq: max eta nsns}
\begin{split}
    \feNSNS & = \frac{8.18 \times 10^{-4}}{0.52 \, \Msun} \times 0.43 \times 0.23 \times 1.0 \\
    & \approx 1.57 \times 10^{-4}
\end{split}
\end{equation}
which is about a factor of 30 higher than the median formation efficiency, but lies very close to the upper $90^{th}$ percentile of the NSNS formation efficiency shown in Figure \ref{fig: yield review}.

Lastly, for BHNS, we assume $M_{\mathrm{ZAMS,1}}>20\Msun$ and $ 8\Msun\leq M_{\mathrm{ZAMS,2}}<20\Msun$. 
Typically, the secondary star will form the NS, and we again adopt $f_{\mathrm{survive \, SN2}} = 1.0$. 
This leads to:
\begin{equation}\label{eq: max eta bhns}
\begin{split}
    \feBHNS & = \frac{4.21 \cdot 10^{-4}}{0.52\Msun}  \times 0.43 \times 1.0 \times 1.0 \\
    & \approx 3.51 \times 10^{-4} \, \Msun^{-1}.
\end{split}
\end{equation}
In this case, the simulations predict values that are significantly lower than our theoretical maximum: Equation \ref{eq: max eta bhns} is seventy times the median and twenty-five times the upper $90^{th}$ percentile of the middle panel in Figure \ref{fig: yield review}.
We further note how the similar values of our maximum \feBHNS and \feBHBH contradict argument 3 from Section \ref{sec:intro}, which posits that the lower formation efficiency of BHNSs is due to their formation with only lower-mass BHs.

The key takeaway from this experiment is that the plateau for BHBH formation at low metallicities is primarily influenced by assumptions about initial distributions, especially the initial mass function (aligning with findings from \citealt{2018A&A...619A..77K}, \citealt{2024arXiv241001451D} and partly with \citealt{2015ApJ...814...58D}).
Surprisingly, the median \feBHBH from simulations at low metallicity is only about eight times lower than the theoretical maximum from eq. \ref{eq: max eta bhbh}.
I.e., many simulations suggest that forming merging BHBH at low $Z$ is quite common, contrary to the belief that it is like `threading the needle'.
\edited{This unexpectedly high efficiency is consistent with recent findings by \cite{2024arXiv241021401S}, who find the fiducial settings in \texttt{SEVN} heavily overestimate the BHBH merger rate in a way that cannot be justified by uncertainties in the metallicity-dependent star formation rate.}
In contrast, the median \feBHNS is nearly two orders of magnitude below a plausible maximum, indicating that complexities of binary interactions have a more significant impact on BHNS formation.
Finally, while the median \feNSNS is about 1.5 dex lower than the theoretical maximum, some models in Figure \ref{fig: yield review} reach this value. 
Generally, we suggest that simulations nearing the calculated maxima might be too optimistic.

\section{A fixed grid of DCO progenitors}\label{sec: methods}
Our back-of-the-envelope calculations from Section \ref{sec: max form eff} can only help us understand what we should expect for the upper limits of double compact object formation. 
However, binary interactions significantly complicate this picture. 
To further investigate why potential double compact object progenitors are lost at different metallicities, we must  simulate a population of binaries.\\

We use the rapid population synthesis code from the {\tt COMPAS} suite\footnote{see also \url{https://compas.science/}} \citep[version \compasv][]{2017NatCo...814906S,2018MNRAS.481.4009V,2022ApJS..258...34R}, to construct a base set of $5 \times 10^6$ binary systems. 
Our fiducial model follows the settings as described in \cite[][]{2022ApJS..258...34R}, with the exception of updated prescriptions for the wind mass loss as described in \tocite{Merritt et al.\ (in prep.)}.
We emphasize that our fiducial model adopts the remnant mass and kick prescriptions outlined in the methods paper: the `delayed' remnant-mass model from \cite{Fryer:2012ApJ}, and a Maxwellian kick distribution with $\sigma_{\mathrm{kick}} = 265 \kms$ \citep{2005MNRAS.360..974H} that is reduced by a fallback fraction for BH natal kicks, while in \compasv, the default settings have changed to use the stochastic remnant masses and corresponding momentum-preserving kicks from \citet{2020MNRAS.499.3214M}. 
We explore a variation with the latter prescriptions to assess their impact on our results. 
Electron-capture, and ultra-stripped SNe are expected to have smaller kicks than core-collapse SNe \citep{2015MNRAS.454.3073S, 2018ApJ...865...61G, 2019MNRAS.484.3307M}.
Hence we adopt $\sigma_{\mathrm{kick}} = 30 \kms$ for these types of SNe.

\begin{figure*}
    \centering
    \includegraphics[width=\textwidth]{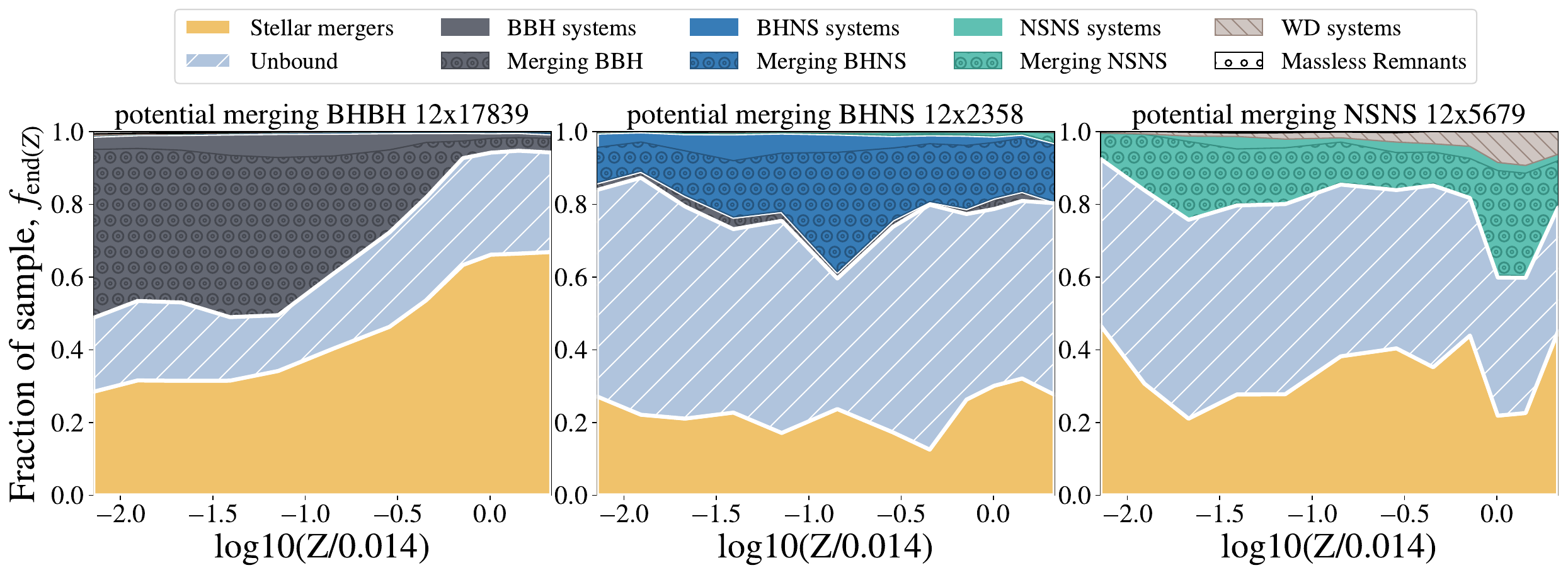}
    \includegraphics[width=\textwidth]{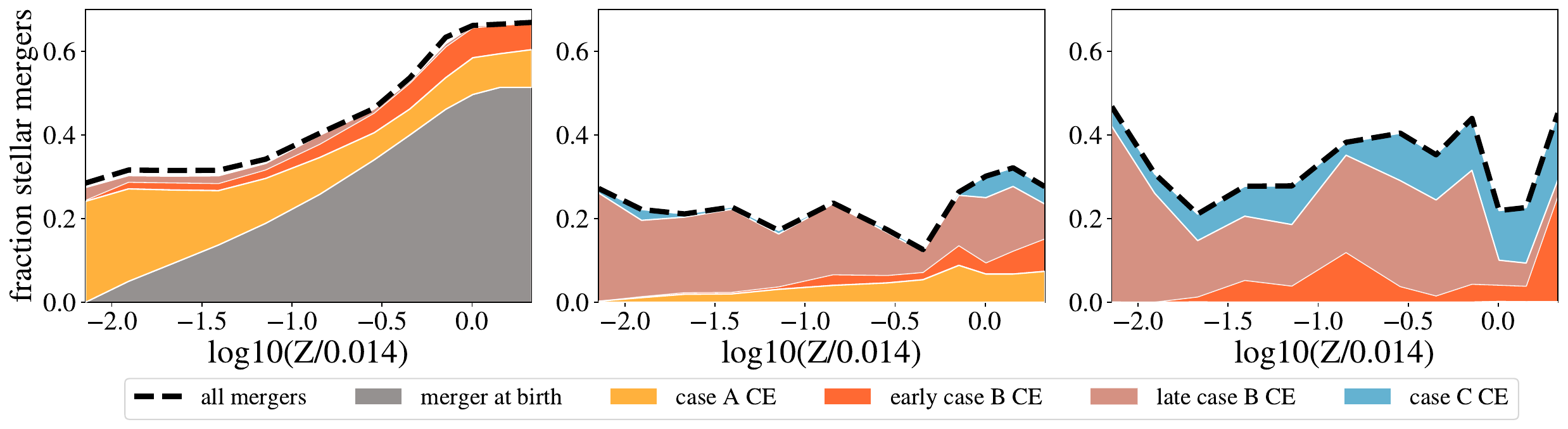}
    
    \caption{\textbf{Top:} The evolutionary end points of potential merging DCO progenitors as a function of metallicity. 
    The strong decrease in BHBH yield at higher metallicity is driven by an increase in stellar mergers and unbound systems, while we find a weak metallicity dependence for merging BHNS and NSNS in this fiducial simulation.
    \textbf{Bottom:} Stellar mergers (yellow area from top panel) colored by the stellar type of the donor star at the onset of the respective mass transfer.
    This shows that, at high metallicities, many potential BHBHs that would have formed through CHE are lost to stellar mergers at birth.     
    \label{fig: N_end_per_Z} }
\end{figure*}

Each system in our base set is characterized by the ZAMS primary mass, mass ratio, period, and the supernova kick magnitude and orientation for both binary components (the latter will only be used when applicable).
To assess the impact of metallicity, we evolve this \textit{exact same set} of binaries at 12 different metallicities.\footnote{metallicities = [$0.0001$, $0.00017321$, $0.0003$, $0.00054772$, $0.001$, $0.002$, $0.004$, $0.00632456$, $0.01$, $0.01414214$, $0.02$, $0.03$]}
These metallicities are chosen to include the metallicities that were used to derive the fitting formulae in \cite{2000MNRAS.315..543H}, on which \COMPAS's single stellar evolution tracks are based. 
To produce the properties of the base set of $5 \times 10^6$ binaries, we draw primary masses between $5\Msun-150\Msun$ from a \cite{2001MNRAS.322..231K} initial mass function, mass ratios from a flat distribution for $q=M_2/M_1~\in~[0.01, 1.0]$ \citep[e.g.,][]{Sana:2012Sci, kobulnicky2014toward}, 
and initial semi-major axes from flat-in-log distribution between $0.01-1000\,$AU. 
After drawing the aforementioned properties once, we use them as seeds at all different metallicities. 
To check that our results are converged, we have repeated our analyses with a reduced resolution of $1\times10^6$ binaries. No significant differences were found. We have furthermore confirmed that the statistical uncertainty is small in our fiducial simulation by bootstrapping the results shown in Figure \ref{fig: N_end_per_Z}. 
All relevant simulation output is publicly available online (see the Software and Data acknowledgment).

\paragraph{Metallicity dependent winds}
Wind mass-loss rates are difficult to model and constrain due to the rarity of massive stars and complications like wind clumping and dust formation \citep[see, e.g., reviews by][]{2014ARA&A..52..487S,2021ARA&A..59..337D,2022ARA&A..60..203V}.
Consequently, both 1D stellar evolutionary and population synthesis codes use simple analytical prescriptions based on limited empirical results. 
Much progress has been made in the last 5 years, and \tocite{Merritt et al.\ (in prep.)} provide a comprehensive overview of updated prescriptions, which form the fiducial model in this work.

In our simulation, we use metallicity-dependent prescriptions for three types of stellar winds:
First, for OB stars, we use main-sequence wind-loss prescriptions from \cite{2021MNRAS.504.2051V}.
This prescription describes a shallower metallicity dependence than \cite{2001A&A...369..574V} of $\dot{M} \propto Z^{0.42}$ for stars hotter than the bi-stability jump, compared to the $\dot{M} \propto Z^{0.85}$ scaling adopted elsewhere.  
Second, for stars with masses above $100\Msun$, we adopt eq. 22 from \cite{2023MNRAS.524.1529S}, which is more appropriate for the expected optically thick winds of such high-mass stars. 
Lastly, for stripped helium or Wolf-Rayet (WR) stars, we apply the prescription from \citet{Vink:2017aap} for low-luminosity stripped stars \citep[see e.g.,][for observational motivation]{2024arXiv240617678R}, and \citet{Sander:2020MNRAS} \citep[including the temperature correction from][]{Sander:2023aap} for high-luminosity WR-like stars.
For more details see \tocite{Merritt et al.\ (in prep.)}.

\paragraph{Red super giant winds}
Although red supergiant (RSG) winds are not inherently metallicity dependent, high rates can cause RSG stars to wind-strip, 
altering interactions with their companions and creating a stripped star (with metallicity dependent winds).
Most studies, including those in Figure \ref{fig: yield review}, apply the wind prescriptions from \citet{deJager:1988aapS} and \citet{Nieuwenhuijzen:1990aap} to all cool stars. 
However, these prescriptions were never intended for RSGs, as the data span the entire HR diagram and include only a hand full of RSGs. 
In this work, we adopt updated empirical mass-loss rates from \citet{2024A&A...681A..17D}, who recalibrated the rates from \citet{Beasor:2020MNRAS} using ALMA observations of RSGs in multiple clusters. These new prescriptions result in mass-loss rates that are orders of magnitude lower than those of \citet{Nieuwenhuijzen:1990aap}, preventing RSGs from wind-stripping their envelopes.

\section{Potential merging DCO progenitors at different metallicities }\label{sec: dominant reason to fail}

From the binary star grid described in Section \ref{sec: methods}, we select all systems that form a merging DCO at \textit{any} of the evaluated metallicities.
\edited{For instance, a system that forms a merging BHBH at one metallicity may instead result in a BHNS or undergo a stellar merger at another metallicity.}
We refer to these systems as ``potential DCOs" and examine their properties across metallicities. 
This helps us understand why a system that becomes a merging DCO at one metallicity may fail to do so at others. 
Unless stated otherwise, a ``merging" DCO is defined as having a coalescence time less than 14\Gyr.

\subsection{The endpoints of potential DCO progenitors}\label{ss: endpoints of DCO}

Figure \ref{fig: N_end_per_Z} shows the evolutionary outcomes of the potential DCOs as a fraction of the total unique number of potential DCOs across metallicities. 
Potential outcomes are grouped into the following classes:
I) A stellar merger occurs, II) The binary becomes unbound (due to a supernova kick), III) A DCO forms i.e., a (merging) BHBH, (merging) BHNS, or (merging) NSNS, IV) One component becomes a white dwarf (WD), or V) A massless remnant.

We see that potential BHBH mergers are lost at higher metallicities, mainly due to an increase in stellar mergers and unbound systems. 
The lack of a large number of wide BHBHs challenges 1C from Section \ref{sec:intro} (which suggested that at high metallicity, stronger stellar winds cause wider separations at DCO formation, resulting in coalescence times that are too long for a merger).
\edited{We further note that massless remnants are not visible in Fig. \ref{fig: N_end_per_Z}. Massless remnants result from pair-instability supernovae, which only occur in the high-mass tail of BH progenitors. As such they account for less than $1\%$ of the potential merging BHBH population at the lowest metallicities.}

The increase in the number of unbound systems at higher metallicity is a wind effect: at higher metallicities, stronger stellar winds result in less massive carbon-oxygen cores, causing larger natal kicks that unbind the binary (1A in Section \ref{sec:intro}). 
This holds true as long as the natal kick is assumed to depend on the progenitor mass.
At the same time, these stellar winds widen the orbit, reducing the orbital velocity and making the system more susceptible to disruption by natal kicks. 
To determine the dominant effect, we additionally created a toy model where we calculate the change in orbital velocity for a stripped star undergoing WR-like winds and the associated kick adjustment due to a lower remnant mass. Assuming the binary is roughly unbound when $v_{\mathrm{kick}} > v_{\mathrm{orb}}$, we find that SN kicks clearly outweigh the effects of orbital widening.
\edited{Although this does not account for the kick direction, we expect this conclusion to hold, as in our models the orbital velocity will decrease by at most a factor of $\sim0.9$, while the kick velocity distribution will range from no kick to a full Maxwell-Boltzmann distribution with $\sigma_{\rm{kick}}=265\kms$.
\footnote{The full calculation can be found at \url{https://github.com/LiekeVanSon/ZdependentFormEff/blob/master/code/AdditionalAnalysisNotebooks/Vorb_vs_Vkick.ipynb} }
}

The number of merging BHNS and NSNS fluctuates with metallicity, but does not show a clear decay with metallicity for our fiducial simulation. 
We furthermore note that the formation yields of BHNS and NSNS are about a factor 8 and 3 lower than the yield of BHBHs respectively (see title of figures), consistent with general findings shown in Figure \ref{fig: yield review}.
The consistently high fraction of unbound systems among potential BHNS and NSNS populations suggests that the formation efficiency of these systems is significantly impacted by natal kicks.

\subsection{Dissecting the stellar merger cases \label{ss: mt cases mergers}}

In the bottom panel of Figure \ref{fig: N_end_per_Z}, we investigate the stellar mergers within the potential DCO progenitors. We divide the stellar mergers by the type of the donor star at the onset of the mass transfer that lead to the merger.
Case A corresponds to main-sequence (MS) donors, early Case B corresponds to mass transfer from a donor on the Hertzsprung gap (i.e., after hydrogen exhaustion, but before the central helium ignition), late case B happens post Hertzsprung gap, but before core helium exhaustion, and Case C corresponds to core helium exhaustion and beyond \citep[e.g.,][]{2008AIPC..990..230D}.

For potential BHBH progenitors, we observe a clear increase in the number of stellar mergers at birth, defined as systems where $R_{1} + R_2 \geq R_{\mathrm{ZAMS}}$. 
The number of case A common envelope mergers decreases at higher metallicity (by about $19\%$ of the potential BHBH). The stellar radius increases at higher metallicity, causing these systems instead to contribute to the stellar mergers at birth at high metallicities. 
The remaining systems that are labeled `stellar mergers at birth' at high metallicities come from stars that evolve chemically homogeneously at low $Z$ but are over-contact systems at high $Z$ (as we will show below in \S \ref{ss: yield per var}). In other words, at low metallicities, these systems successfully form BHBH mergers through the CHE channel. However, at higher metallicity, the larger radii of these systems no longer fit within the orbit, causing them to `merge at birth'.
One could argue about the significance of systems that `merge at birth,' as they were technically never born as binaries. 
However, the key takeaway here is that, under our fiducial assumptions, the window for certain very tight orbit formation channels, like the CHE channel, closes at higher metallicities.

We furthermore observe a modest rise in the number of early Case B common envelope mergers at higher metallicity. This is also a radius effect, as it reflects an increased interaction space during Hertzsprung gap phase. Common envelopes that were case C at low metallicity, become early case B at high metallicity. 
\edited{Under our default ``pessimistic common envelope" assumption, all common envelopes from donors with presumably radiative envelopes lead to stellar mergers (i.e., HG donors, see also argument 4 in Section \ref{sec:intro}). 
If we had chosen the ``optimistic" common envelope assumption, the extra early case B mergers at high metallicity would not occur, highlighting the importance of common envelope ejection assumptions.
Nonetheless, this effect appears far from dominant. }
There is a similar increase in Case A and early Case B common envelope mergers in the potential BHNS and NSNS systems at higher metallicities.
However, the total number of stellar mergers for the BHNS and NSNS progenitors generally fluctuates heavily, and does not show any clear trend with metallicity. \\

We conclude that in our fiducial model, the primary cause of potential BHBH mergers being lost at high metallicity is an increase in stellar mergers. This rise in stellar mergers is largely due to an increase in mergers occurring at birth, and, to a lesser extent, a higher number of early case B common envelope events (argument 2 from Section \ref{sec:intro}).
Since this conclusion relies heavily on the uncertain contribution of the CHE channel to BHBH formation, we continue our investigation per formation channel in the sections below. 
The second leading factor for the loss of potential BHBH systems at high metallicity is their being unbound by natal kicks (arguments 1A and 1B from Section \ref{sec:intro}, as we do not distinguish between unbound BHBH and BHNS).
In our fiducial model, both the potential BHNS and NSNS systems do not show a significant trend with metallicity. We do observe a modest increase in the number of early case B common envelopes leading to merger at higher metallicities, but this does not increase the total number of stellar mergers.

\section{Assessing robustness across assumptions} \label{sec: phys var}

\begin{figure*}
    \centering
    \includegraphics[width=0.9\textwidth]{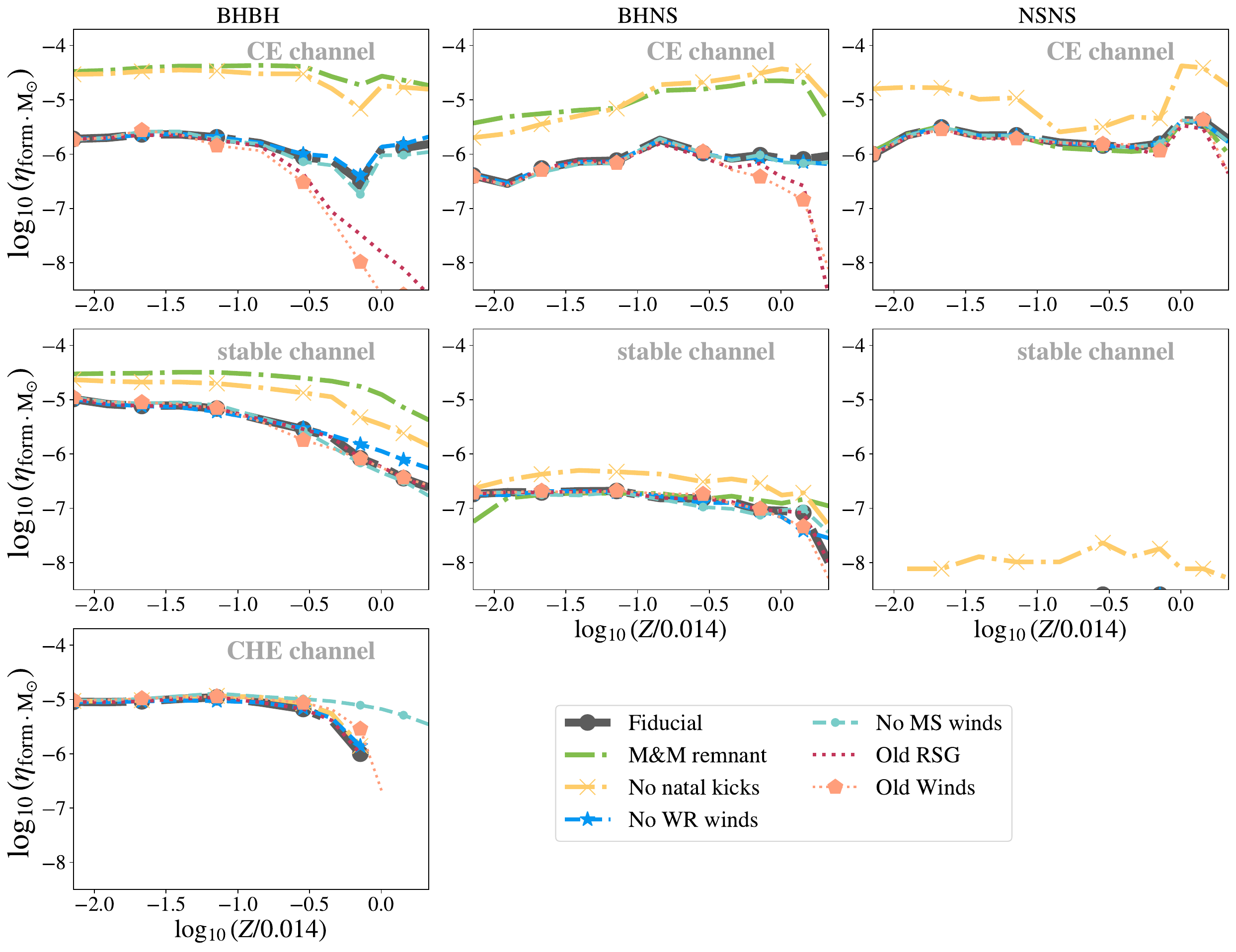}
    \caption{The formation efficiency split by DCO flavour and formation channel. We distinguish between the chemically homogeneous evolution (CHE), only stable mass transfer, and common envelope (CE) channels.
    All three formation channels contribute to BHBH formation, while only stable mass transfer and CE contribute to BHNS formation, and NSNS are dominated by the CE channel. The metallicity dependence of the dominant formation channel will determine the metallicity dependence of the DCO flavour. 
}
    \label{fig: yield per var}
\end{figure*}

To assess the robustness of our findings in Section \ref{sec: dominant reason to fail}, we run several variations of our fiducial simulation. 
Specifically: i) turning off chemically homogeneous evolution (no CHE), ii) applying the remnant mass and natal kick prescriptions from \cite{2020MNRAS.499.3214M} (M\&M remnants), iii) excluding BH kicks (No BH kicks), iv) excluding both BH and neutron star (NS) natal kicks (No natal kicks), v) removing Wolf-Rayet winds (No WR winds), vi) removing main-sequence and very-massive-star winds (No MS winds), vii) excluding all stellar winds (No winds), viii)  using the RSG wind prescriptions from \cite{Nieuwenhuijzen:1990aap} (Old RSG), ix) using all the wind prescriptions from \cite{2022ApJS..258...34R} (Old Winds).

\subsection{Formation efficiency per channel and physics variation \label{ss: yield per var}}
In Figure \ref{fig: yield per var} we show the formation efficiency for a selection of these variations. 
The formation efficiencies are broken down by formation channel for each DCO type: systems that experienced at least one common envelope (CE, top row), those that underwent only stable mass transfer (middle row), and those formed through chemically homogeneous evolution (CHE, bottom row) \citep[see, e.g.,][for a review of these formation channels]{2022PhR...955....1M}.

The first thing we note is that while all three formation channels contribute to BHBH formation, only the stable mass transfer and CE channel contribute to BHNS formation, and NSNS systems are almost exclusively formed through the CE channel.
This plays an important role in understanding why BHBH formation is metallicity-dependent, while NSNS formation is not.

BHBH mergers are shown in the left-most column. 
In our fiducial simulation, the CHE channel dominates the formation efficiency of BHBH mergers with an almost flat value of $\log_{10}(\feBHBH\Msun)=-5$ until $\log_{10}\left( Z / Z_\odot \right) \simeq -0.5$.
Indeed, the CHE channel is the dominant contributor to the yield in all variations, except for those with weaker BH kicks (No Natal Kicks, and M\&M remnants).
The CHE channel typically shows a sharp decline above $\log(Z/0.014) \approx -0.25$, except for in the No MS Winds variation, where the CHE channel maintains a high formation efficiency at high metallicities. This is because strong MS winds widen the orbits, thereby reducing their tidally locked spins, and so prevent them from evolving chemically homogeneously \citep[cf.][]{2021MNRAS.505..663R}.  
When CHE is excluded, the stable MT channel dominates the formation efficiency at metallicities $\log_{10}(Z)\lesssim-0.5$, for most variations (again with the exception of No Natal kicks and M\&M). 
It displays a moderately flat formation efficiency below $\log_{10} \left( Z /\Zsun \right) \simeq -1.0$, followed by a somewhat shallow decline of about 1-1.5 dex.
\edited{We note that the M\&M variation yields a higher BHBH formation efficiency compared to the no natal kick prescription. This is because, in addition to weak BH natal kicks, the M\&M remnant mass prescription produces more BHs at the expense of NSs than the \cite{Fryer:2012ApJ} delayed prescription.}

The CE channel shows relatively flat behavior with respect to metallicity in most variations, except for the Old Wind and Old RSG variations, where the formation efficiency declines steeply at high metallicities. 
This decline occurs because the `old' RSG wind prescriptions from \citet{Nieuwenhuijzen:1990aap} are orders of magnitude stronger than the newer values from \citet{2024A&A...681A..17D}. At high metallicities, stars are less luminous and thus may avoid the Luminous Blue Variable (LBV) regime \citep{2020Galax...8...20W}. With the less strong \cite{2024A&A...681A..17D,Beasor:2023MNRAS} winds, these stars retain their envelopes and undergo common envelope evolution. On the other hand, with the older \citet{Nieuwenhuijzen:1990aap} prescription, these stars self-strip at high Z, avoiding common envelope evolution. 
This primarily affects a sub-channel of the CE channel where the more massive star initiates a common envelope during the first mass transfer phase. 

For BHNS mergers, the CE channel dominates the stable MT formation efficiency in all explored variations.
Both the stable MT and CE channels generally show flat behavior with respect to metallicity across most variations.
The exceptions are the Old RSG and Old wind variations, which decline at high metallicities, similar to the BHBH merger case. 
These winds affect stars at the edge of the LBV regime ($20-30 \Msun$ at ZAMS), which mainly form low-mass BHBH systems. The CE channel primarily produces lower-mass BHs \citep[see \S3.1 in][]{2022ApJ...931...17V}, which explains the strong impact.
Both the M\&M remnants and No Natal Kicks variations furthermore show an increase in formation efficiency with metallicity. 
The M\&M remnants prescription applies much weaker BH kicks than our fiducial model using the Fryer delayed prescription.
This indicates that BH kicks, rather than NS kicks, play a more crucial role in BHNS formation, as the BH usually forms first, at a time when the binary is wider and more susceptible to disruption. 

NSNS formation is almost exclusively driven by the CE channel \citep[in line with earlier results][]{2022ApJ...937..118W,2023MNRAS.524..426I,2023ApJ...955..133G}. 
The stable mass transfer channel is unsuccessful when low ZAMS mass stars are involved because the extreme mass ratios at the onset of mass transfer from the secondary star lead to unstable mass transfer instead \citep[as described in more detail in][]{2022ApJ...940..184V,2024A&A...681A..31P}. 
Most variations have little impact on efficiency, except for the No Natal Kicks variation, which significantly increases it.

In summary, Figure \ref{fig: yield per var} highlights three key points: (I) the role of metallicity in \fe depends on the formation channel, (II) BHBH formation appears less sensitive to metallicity than previously thought, with \feBHBH decreasing by several dex at high $Z$, only when Old RSG winds are applied, III) NSNS only form through the CE channel, making it particularly relevant whether this channel is metallicity dependent or not.

\subsection{Summary endpoints per variation}\label{ss: summary endpoints}

\begin{figure*}
    \centering
    \includegraphics[width=0.95\textwidth]{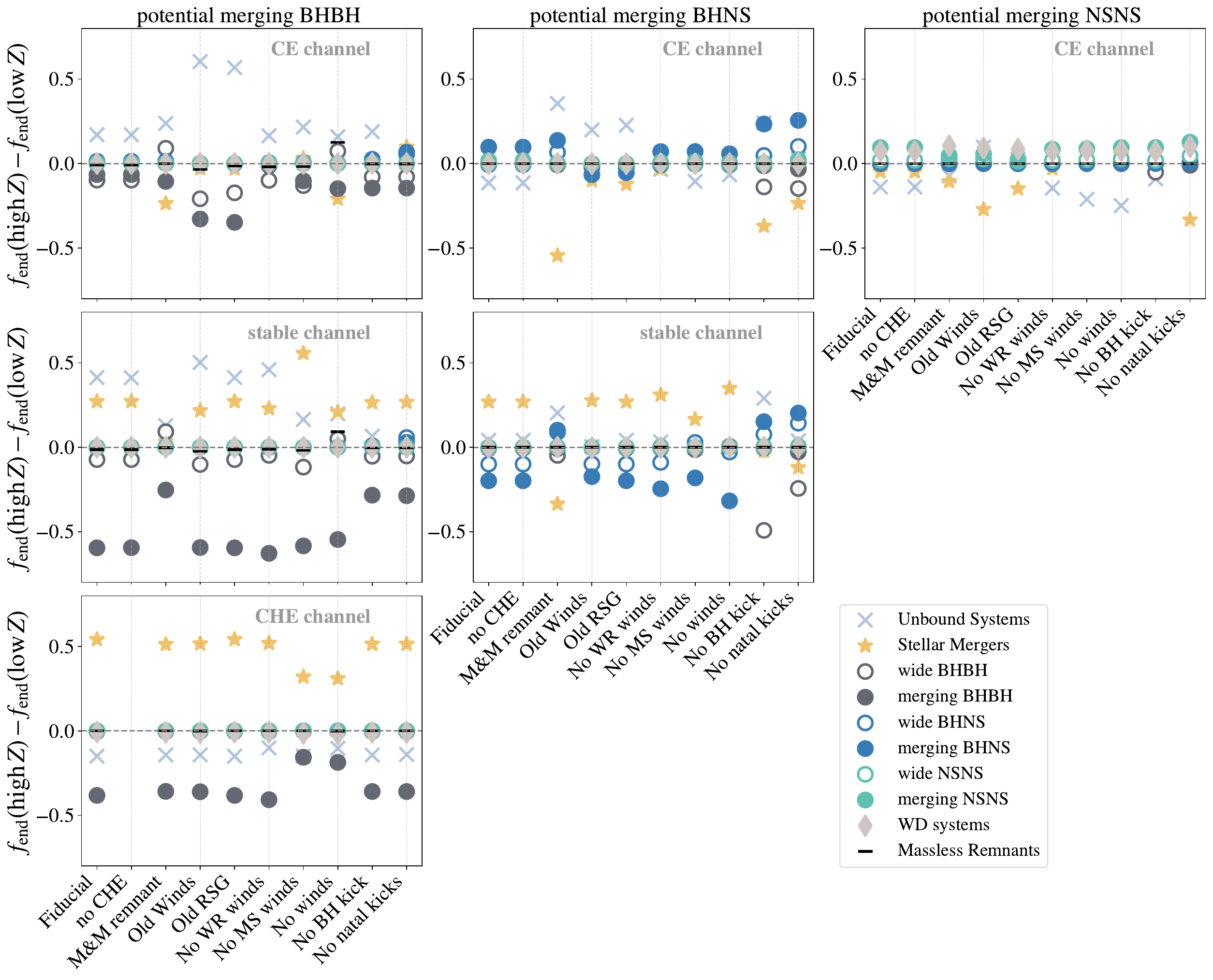}
    \caption{The net change in evolutionary endpoints (as in Figure \ref{fig: N_end_per_Z}), for all simulated variations (x-axes) and formation channels (columns). 
    Endpoints that lie below the null line are more frequent at low metallicity and less frequent at high metallicity, where they are replaced by (the sum of) endpoints above the null line. 
    This shows that the metallicity dependence of each DCO type is a question of formation channel.
    }
    \label{fig: summary end evol}
\end{figure*}

To identify the dominant physics responsible for any metallicity dependence in the formation efficiency, we analyze the evolutionary endpoints $f_{\mathrm{end}}(Z)$ as in the top panel of Figure \ref{fig: N_end_per_Z}, for each variation. 
We summarize the net change between low and high metallicity in Figure \ref{fig: summary end evol}.
That is, we subtract the average $f_{\mathrm{end}}(Z)$ at the lowest two metallicities from that at the highest two metallicities:
\begin{equation}
        f_{\mathrm{end}}(\mathrm{high} \, Z) -f_{\mathrm{end}}(\mathrm{low} \, Z),
\label{eq: f_end high low Z}
\end{equation}
where 
\begin{align}
    f_{\mathrm{end}}(\mathrm{high} \, Z) & = \frac{f_{\mathrm{end}}(0.03) + f_{\mathrm{end}}(0.02) }{2}  \\
    f_{\mathrm{end}}(\mathrm{low} \, Z)  & =  \frac{f_{\mathrm{end}}(1\cdot 10^{-4}) + f_{\mathrm{end}}(1.7\cdot 10^{-4}) }{2} .
\end{align}

We calculate $f_{\mathrm{end}}(Z)$ for each possible endpoint (such as stellar mergers, unbound systems, BHBH, etc.), for every variation, and every formation channel. For example, we select all systems that form a DCO merger through the CE channel at any metallicity (top row of Figure \ref{fig: summary end evol}). If a system forms a merging DCO via e.g., the CE channel at one metallicity but via e.g., the stable channel at another, it will appear in both categories.
Figure \ref{fig: summary end evol} should be interpreted as the change in dominant endpoints between low and high metallicity. 
\edited{In other words, endpoints below the null line are more prevalent at low metallicity, while endpoints above the null line are more common at high metallicity.}
If the points lie around the null line, then there is little difference between low and high Z.
We note that the metallicity dependence of each DCO flavor is clearly dictated by whichever formation channel dominates their formation. 
The metallicity dependence (or lack thereof) of each formation channel is driven by different physics and therefore sensitive to different variations. 

\paragraph{The CHE channel} is only effective for BHBH formation (bottom left panel of Figure \ref{fig: summary end evol}), and loses potential merging BHBHs to stellar mergers at high metallicity. This is a radius-driven effect: stars that undergo CHE at low metallicity have larger radii at higher metallicities, leading to stellar mergers at the ZAMS. 
This explains the significant increase in stellar mergers at high metallicity shown in the top panel of Figure \ref{fig: N_end_per_Z}.

\paragraph{The stable MT channel} is effective for both BHBH and BHNS formation (middle row of Figure \ref{fig: summary end evol}). 
The formation efficiency of BHNS \textit{increases} with metallicity when natal kicks are low or nonexistent (no BH kick, no natal kick and M\&M remnants variations).  
The BHNS formation efficiency decreases with metallicity across all other variations, where potential progenitors are lost to stellar mergers.
Stellar mergers also cause a significant loss in potential BHBH in nearly all variations.
For both BHBH and BHNS, the loss of potential DCO progenitors to stellar mergers is a metallicity-dependent radial expansion effect: the larger radii cause mass transfer to occur earlier in the evolution. 
For BHNS, we find that the increased number of stellar mergers comes from I) an increase in first mass transfer case A common envelopes II) an increase in early case B common envelopes from star 2, both at the expense of stable early case B mass transfer. 
For BHBH we find an increase in stable case A mass transfer from the first star at the expense of stable early case B. The more conservative nature of case A mass transfer results in skewed mass ratios at the onset of mass transfer from the second star, leading to more common envelopes at the expense of the stable mass transfer channel.
\edited{Similarly, one might expect an increase in early case B mass transfer at the expense of late case B or case C mass transfer. However, this does not occur because stable mass transfer does not sufficiently shrink orbits wider than about $a_{\mathrm{zams}} \gtrsim 200 \Rsun$ to produce a GW merger, thereby excluding the contribution from late case B donors.}
Potential BHBH systems are furthermore lost at higher metallicities to unbound systems in the fiducial, Old Winds, Old RSG winds and no WR wind variations. 
We find that this is again largely driven by a radial expansion: stable early case B mass transfer becomes stable case A mass transfer.
Case A mass transfer leads to much lower-mass cores in COMPAS (and most BSE-based codes), causing more BH progenitors to be unbound by the first SN \citep[cf.][]{2024arXiv240900460S}.
The M\&M remnants variation loses potentially merging BHBH at high metallicity to unbound systems, and to BHBH systems that are too wide to lead to GW emission and hence a DCO merger.

\paragraph{The common envelope channel} (top row of Figure \ref{fig: summary end evol}) is an effective formation channel for all three DCO flavors. We find it to be generally less sensitive to metallicity with respect to the other two channels (the end points are more often centered around the null line). 
At high metallicities, potential BHBH systems are lost to unbound systems.
This occurs because metallicity-dependent winds reduce core masses. Since BH natal kicks are assumed to be reduced by fallback, smaller core masses result in larger natal kicks, while NS progenitors experience the full strength of their natal kick.
The largest change with metallicity is in the Old Winds and Old RSG Winds variations. As discussed in \S \ref{ss: yield per var}, the old RSG winds can additionally strip stars at high $Z$, preventing the onset of a CE. 
The No BH Kick variation still shows a significant increase in unbound systems at high $Z$, suggesting that these systems contain a NS whose natal kick leads to unbinding.
In the No Natal Kick variation, potential BHBH systems evolve into BHNS systems instead of unbound systems. 
In essentially all variations the formation rates of BHNS and NSNS are somewhat higher in the highest metallicity bins with respect to the lowest metallicity bins. 
We do not observe an increase in unbound systems with metallicity in the pool of potential NSNS systems, as NSs are always assumed to receive a natal kick. 
Instead, we consistently see an increase in systems containing a WD. 
These WD + NS systems are an interesting target for the LISA mission \citep{2017arXiv170200786A}. 

\subsection{Evidence for mass-dependent BH kicks} \label{ss: obs BH kicks}

In \S \ref{ss: yield per var} and \S \ref{ss: summary endpoints}, we found that potential BHBH mergers are predominantly lost to unbound systems in the CE channel at high metallicity. 
This increase in unbound systems is driven by two processes in our simulations: first, the metallicity dependent winds are stronger at high $Z$, lowering the core masses. Second, BH kicks are reduced by fallback, meaning that smaller core masses receive larger natal kicks. 
This particularly affects the CE channel, as it primarily produces lower-mass BHs \citep[\S3.1 in][]{2022ApJ...931...17V}, and is thus dominated by systems near the threshold for receiving a natal kick. 
While the exact scaling of radiation-driven winds with metallicity remains debatable, the metallicity dependence of these winds is well established \citep{1987A&A...173..293K,2007A&A...473..603M,2015A&A...581A..21H,2020MNRAS.491.4406S}. 
The process of natal kicks and fallback onto proto-BHs is far less understood.

Theoretically, which stars leave behind NSs and which BHs is still unresolved. 
This question is also tied to the putative `NS-BH mass gap/dearth' in the $3\text{--}5\mathrm{M}_\odot$ range \citep{1997AAS...190.1001B,2001ApJ...554..548F,2010ApJ...725.1918O,2011ApJ...741..103F}.
Recent discoveries like GW230529 \citep{2024arXiv240404248T} and the low-mass system in Gaia DR3 \citep[UCAC4 569-026385][]{2024NatAs.tmp..215W}, suggest that at least some BHs are born with low masses.  
The natal-kicks of these systems are even more uncertain. Some studies indicate that the kick velocity for BH-forming collapses could be only a few km/s \citep{2020PhRvD.101l3013W,2024Ap&SS.369...80J}, while others suggest that even $40\Msun$ ZAMS stars might receive kicks of several hundred km/s \citep[e.g.,][]{2022MNRAS.517.3938C,2023ApJ...957...68B}. 
Additionally, we lack a clear understanding of the fallback process during and post explosion \citep[e.g.,][]{2018ApJ...862L...3S,2020MNRAS.495.3751C}.

\begin{figure}
    \includegraphics[width=0.475\textwidth]{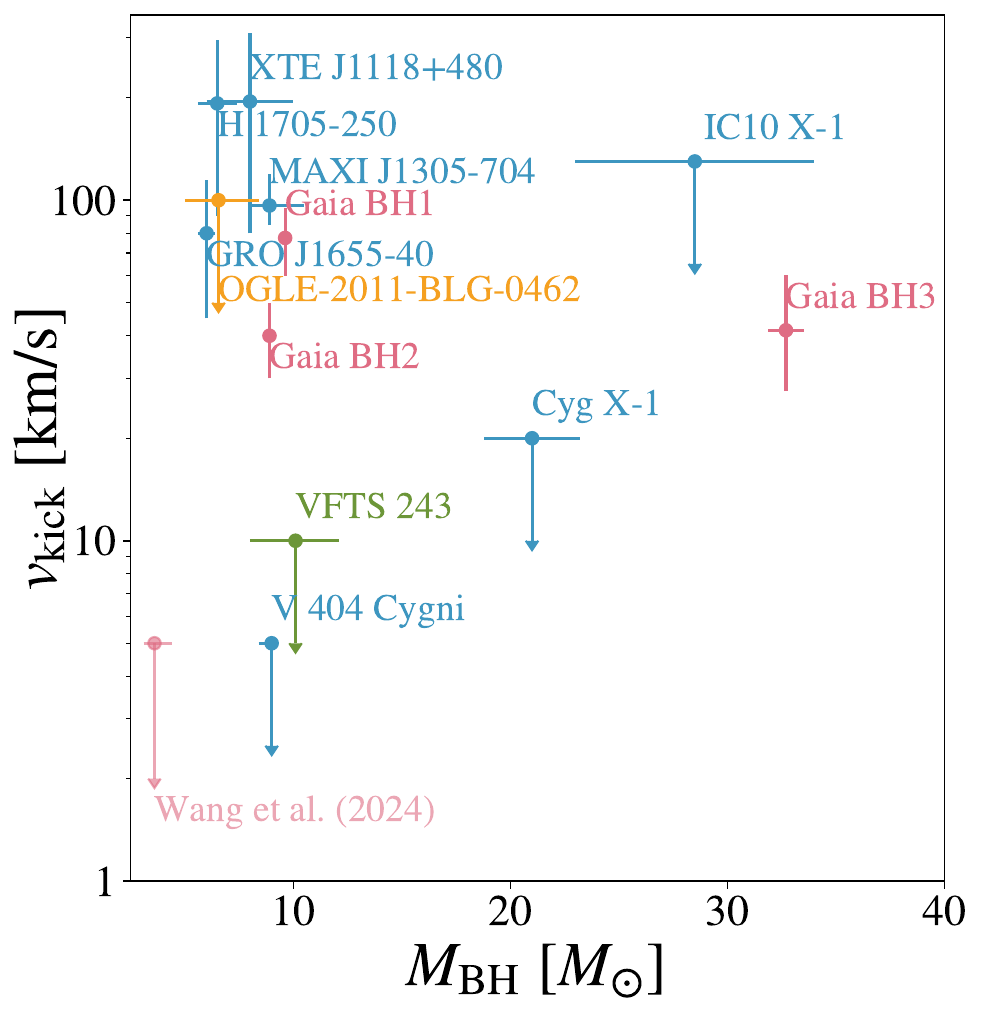}
    \caption{Observational estimates of BH natal kick velocities as a function of black hole mass. The colors indicate the method with which it was first detected: XRB (blue) microlensing (yellow), spectroscopy (green) or astrometry (pink).
    These observations do not yet place any meaningful constraint on a relationship between BH mass and its natal kick. }
    \label{fig: BH kicks}
\end{figure}

Observational constraints for BH natal kicks are quite limited. 
Currently, only a handful of systems have well-constrained BH natal kicks. 
We review most known observational constraints of individual systems in Figure~\ref{fig: BH kicks}.
In blue we show mass and inferred constraints on the natal kicks of the X-Ray binaries GRO J1655-40 \citep{2016A&A...587A..61C, 2005ApJ...625..324W}, XTE J1118+480 \citep{2009ApJ...697.1057F}, MAXI J1305-704  \citep{2021MNRAS.506..581M, 2023ApJ...952L..34K}, V 404 Cygni \citep{2010ApJ...716.1105K, 2024arXiv240403719B}, H 1705-250 \citep{2024MNRAS.527L..82D}, IC10 X-1 \citep{2014ApJ...790..119W}, and Cyg X-1 \citep{2021Sci...371.1046M,2014ApJ...790..119W}.
We furthermore show the spectroscopically detected VFTS 243 (\citealt{2022NatAs...6.1085S,2024PhRvL.132s1403V}, see also \citealt{2023ApJ...959..106B}) in green, and the microlensed system OGLE-2011-BLG-0462 \citep{2022ApJ...930..159A, 2022ApJ...933...83S, 2023ApJ...955..116L} in yellow. 
Lastly we add the astrometrically detected Gaia BH1 \citep{2023MNRAS.518.1057E}, Gaia BH2 \citep{2023MNRAS.521.4323E, 2024arXiv240313579K}, and Gaia BH3 \citep{2024A&A...686L...2G} in pink. For the Gaia BHs we show the kick velocity under the assumption that these systems formed from isolated binaries following \cite{2024arXiv240313579K} for BH1 and BH2, and the $68\%$ range from \cite{2024arXiv240413047E} for BH3.
However, we stress that Gaia BH3 likely formed through dynamical assembly \citep[e.g.,][]{2024A&A...687L...3B,2024A&A...688L...2M,2024arXiv240413047E}. 
Although we have no formal constraint on the natal kick of Gaia DR3 3425577610762832384 \citep[][]{2024NatAs.tmp..215W} its surprisingly wide and circular orbit of 880 days suggests it was born with a natal kick of at most a few \kms. 

\edited{
We conclude that these observations do not yet provide meaningful constraints on the relationship between BH mass and natal kicks. The data could support a mass-dependent trend, but are equally consistent with a bimodal distribution independent of mass. 
Interestingly, recent work by \citet{2024arXiv241116847N} finds no convincing evidence for mass-dependent kicks within the BH population, though their data hints at a bimodal distribution, similar to the kick distribution observed for NS \citep[in agreement with][]{2023MNRAS.525.1498Z}.
As shown in this work, if BH natal kicks are not dependent on their mass, this would have significant implications for the metallicity dependence of BHBH formation efficiency.
This underscores the urgency of constraining mass–natal kick relations through observations.}

\section{Summary \label{sec: conclusion}}

In this work, we investigate the remarkably consistent predictions for the metallicity dependence of various DCO formation efficiencies (see Figure \ref{fig: yield review}). 

We first demonstrate how the yield in the low-metallicity plateau in BHBH formation efficiency is set by the assumed mass and period distributions at birth (Section \ref{sec: max form eff}). Notably, this plateau implies that if BHs do not receive a natal kick, the efficiency of merging BHBH formation is quite high: approximately \textit{one in every eight} binary systems with masses high enough to form a BHBH, and separations small enough to interact, will lead to a merging BHBH. 
Further metallicity dependence of DCO formation (or lack thereof) is a channel dependent question: 

\begin{itemize}[nolistsep,leftmargin=1em]
    \item  The formation efficiency of the CE channel with metallicity is generally flat (Figure \ref{fig: yield per var}). 
    We only find a very steep (multiple orders of magnitude) drop in BHBH formation efficiency when Old RSG wind prescriptions \citep{Nieuwenhuijzen:1990aap} are used. 
    These winds strip stars of their envelopes (thereby evading CE) even when they avoid the LBV regime at high metallicity. Newer prescriptions \citep{Beasor:2023MNRAS,2024A&A...681A..17D} prevent self-stripping, causing the flatter metallicity dependence we observe.
    The CE channel further loses potential BHBH progenitors to unbound systems at high metallicities (Figure \ref{fig: summary end evol}). 
    This is a wind-dependent effect: stronger winds lead to lower core masses, and thus stronger kicks which unbind the system. 
    For potential NSNS stars this effect does not exist because NSs are assumed to receive a natal kick that is not proportional to their progenitor mass. 
    
    \item The stable mass transfer channel loses BHBH and BHNS progenitors at higher metallicity because stellar expansion, and hence mass transfer, happen earlier in the donor's evolution (see Figure \ref{fig: SSE properties}). 
    This leads to an increase in case A mass transfer, at the expense of stable early case B mass transfer. 
    This decreases the contribution of the stable mass transfer channel because it either skews the mass ratio at the onset of the second mass-transfer phase, leading to unstable mass transfer, or lowers core masses, increasing natal kicks and unbound systems (though this may reflect shortcomings in our handling of case A mass transfer). 
    Similar metallicity-dependent radius effects impact NS progenitors, but NSNS mergers do not form through the stable mass transfer channel due to constraints imposed by the critical mass ratio for stability \citep[cf.][]{2022ApJ...940..184V}.

    \item The CHE channel is only effective for BHBH formation. 
    The window for CHE is $Z$ dependent, and bounded by ZAMS radii. Higher metallicity stars have larger radii, leading to stellar mergers at ZAMS (bottom panel of Figure \ref{fig: N_end_per_Z}).
    At larger periods, the window closes due to reduced stellar rotation rates, with main sequence winds significantly widening the orbit at high metallicities.   
\end{itemize}

We conclude that BHBH formation may be less dependent on metallicity than previously thought. 
Key factors influencing DCO formation efficiency and its metallicity dependence are:
1) The initial distributions of masses and binary periods \citep[see also][]{2024arXiv241001451D}.
2) Conditions for CHE, including the radii of massive stars on the zero-age main sequence (ZAMS) at varying metallicities. 
3) The radial evolution of stars at different metallicities, and the corresponding progression of stable mass transfer, specifically mass transfer efficiency and stability. 
4) The uncertain mass dependence of natal kicks (see \S\ref{ss: obs BH kicks}). Additional electromagnetic observations of BHs will be particularly valuable in constraining these uncertainties, such as Gaia astrometry \citep[e.g.,][]{2024OJAp....7E..58E}, microlensing studies \citep[e.g.,][]{2023hst..prop17400L} and detached BH systems like VFTS 243 \citep[e.g.,][]{2022NatAs...6.1085S}.

The formation efficiency of merging DCOs with metallicity can furthermore be inferred from GW observations. Evidence from GWTC-3 already suggests higher BHBH formation efficiency in low-metallicity environments \citep{2024ApJ...970..128S}, though these results remain heavily model-dependent due to limited data. 
As the GW-source catalog grows, particularly with the advent of third-generation observatories like the Einstein Telescope \citep{2020JCAP...03..050M} and Cosmic Explorer \citep{2021arXiv210909882E,2023arXiv230613745E}, we expect the largest increase in the detection of lower-mass DCOs.
These observations will provide critical insights into the nuances of different formation channels and will be able to confirm or disprove their metallicity dependence as described in this work.

\begin{acknowledgments}
The authors would like to thank S. Stevenson and J. Riley for their essential contributions that facilitated the variations and updates in the wind prescriptions.
LvS also thanks J. Goldberg for numerous constructive conversations that contributed to the development of this work. 
%
SKR thanks the Center for Computational Astrophysics at the Flatiron Institute for hospitality while a part of this research was carried out. The Center for Computational Astrophysics at the Flatiron Institute is supported by the Simons Foundation.
IM acknowledges support from the Australian Research Council (ARC) Centre of Excellence for GravitationalWave Discovery (OzGrav), through project number CE230100016.
AACS is supported by the German Deutsche Forschungsgemeinschaft (DFG) under Project-ID 445674056 (Emmy Noether Research Group SA4064/1-1, PI Sander).
\end{acknowledgments}

\section*{Software and Data \label{sec: software data}}
All code associated to reproduce the data and plots in this paper is publicly available at \url{https://github.com/LiekeVanSon/ZdependentFormEff}. 
This study makes use of rapid population synthesis code from the {\tt COMPAS} suite version \compasv \citep[][]{2017NatCo...814906S,2018MNRAS.481.4009V,2022ApJS..258...34R}.
The data used in this work is available on Zenodo under an open-source Creative Commons Attribution license at
\dataset[10.5281/zenodo.13999532]{https://doi.org/10.5281/zenodo.13999532}.
Further software used in this work: Python \citep{PythonReferenceManual},  Astropy \citep{astropy:2013,astropy:2018,astropy:2022} Matplotlib \citep{2007CSE.....9...90H},  {NumPy} \citep{2020NumPy-Array}, SciPy \citep{2020SciPy-NMeth}, \texttt{ipython$/$jupyter} \citep{2007CSE.....9c..21P, Kluyver2016jupyter},  Seaborn \citep{waskom2020seaborn},  {hdf5}   \citep{collette_python_hdf5_2019}, and WebPlotDigitizer \citep{WebPlotDigitizer}.

\appendix

\section{Calculating the Drake equation}\label{app: drake}
The calculations in this appendix can also be found here: \url{https://github.com/LiekeVanSon/ZdependentFormEff/blob/master/code/analytical_yield.ipynb}.
We approximate the terms within the round brackets of Equation \ref{eq: drake} 
with the probability to form a pair of massive stars with the `right' set of (initial) conditions. The conditions in questions are random variables at ZAMS (the primary mass, $m_1$, secondary mass, $m_2$, orbital separation, $a$), and factors affecting the stars' survival during supernovae in the first and second mass transfer phases. The joint probability density function (pdf) describing occurrence of a specific type of compact binary merger is given by:

\begin{equation}\label{eq:pdf}
\begin{split}
p(m_1, ~m_2, ~a, ~\text{survive SN1}, ~\text{survive SN2}) 
& = p(m_1) \times p(m_2~|~m_1) \\
& \times p(a ~|~m_1, ~m_2) \\
& \times  p(\text{survive SN1}~|~m_1, ~m_2, ~a) \\
& \times p(\text{survive SN2}~|~m_1, ~m_2, ~a, ~\text{survive SN1}) \\
& \approx p(m_1) \times p(m_2|m_1) \times p(a) \times p(\text{survive SN1}) \times p(\text{survive SN2})
\end{split}
\end{equation}

By integrating over the relevant ranges of $m_1, ~m_2, ~a, ~\text{survive SN1}, ~\text{survive SN2}$, we obtain the probability for a specific type of compact binary merger to occur. 

\subsection{Primary and secondary masses}
We aim to compute the probability that $m_2$ falls within a certain mass range $[c,d]$ given that $m_1$ within a certain mass range $[a,b]$. We can write $m_2$ in terms of the mass ratio $q \equiv m_2/m_1 < 1$:
\begin{equation} \label{eq: pm2 in q}
    p(q~|~m_1) = U\left(q \mid \frac{0.01 \, \Msun}{m_1}, 1\right), 
\end{equation}

\edited{where $U$ denotes a uniform distribution. }

We assume $p(m_1)$ follows the Kroupa IMF, defined as:
\begin{equation} \label{eq: kroupa}
\text{Kroupa IMF}(m_1) = C_1
    \begin{cases}
      (m_1/\Msun)^{-0.3} & 0.01~\Msun \leq m_1 < 0.08~\Msun\\
      C_2 (m_1/\Msun)^{-1.3} & 0.08~\Msun \leq m_1 < 0.5~\Msun\\
      C_3 (m_1/\Msun)^{-2.3} & 0.5~\Msun \leq m_1 < 300~\Msun\\
      0 & \text{otherwise}
    \end{cases}       
\end{equation}

Here $C_2 = 0.08^{-0.3 + 1.3} = 0.08$ and $C_3 = C_2 \cdot 0.5^{-1.3 + 2.3} = 0.8 \cdot 0.5$ to ensure continuity. 
The constant, $C_1$ is determined by normalizing the IMF over the entire mass range:
\begin{equation*}
    C_1 = \Big[ \frac{0.08^{(-0.3+1)}-0.01^{(-0.3+1)}}{(-0.3+1)} + 0.08 \left(\frac{0.5^{(-1.3+1)}-0.08^{(-1.3+1)}}{(-1.3+1)} \right) + 0.08 \cdot 0.5 \cdot \left(\frac{300^{(-2.3+1)}-0.5^{(-2.3+1)}}{(-2.3+1)} \right) \Big]^{-1}  \, \Msun^{-1}
\end{equation*}

Note that we set the minimum and maximum stellar masses to $0.01~\Msun$ and $300~\Msun$, respectively.

Hence $f_{\text{primary}} \times f_{\text{secondary}}$ when $0.5 \, \Msun < a < m_1 < b < 300 \, \Msun$ and $0.01 \, \Msun < c < m_2 < \min\left( m_1, d \right) < 300 \, \Msun$ is given by:

\begin{equation}\label{eq:pdf_BBH}
    \begin{split}
        f_{\text{primary}} \times f_{\text{secondary}} 
        & = \int_{m_1=a}^{m_1=b} dm_1  \int_{q=c/m_1}^{q=\min\left(1, d/m_1\right)} dq  ~ p(m_1) \times p(q|m_1) \\
        & = \int_{m_1=a}^{m_1=b} dm_1 \int_{q=c/m_1}^{q=\min\left(1, d/m_1\right)} dq ~\text{Kroupa IMF}(m_1) \times U(q|\frac{0.01\Msun}{m_1},1) \\
        & = \int_{m_1=a}^{m_1=b} dm_1 ~\text{Kroupa IMF}(m_1) \frac{\min\left(1, \frac{d}{m_1}\right) - \frac{c}{m_1}}{1 - \frac{0.01 \, \Msun}{m_1}} \\
        & = \int_{m_1=a}^{m_1=b} dm_1 ~\text{Kroupa IMF}(m_1) \times  \left( \frac{\min\left(m_1, d\right) - c}{m_1 - 0.01 \, \Msun} \right)  \\
        & = \int_{m_1=a}^{m_1=b} dm_1 C_1 \cdot C_3 \cdot  (m_1/\Msun)^{-2.3} \times  \left( \frac{\min\left(m_1, d\right) - c}{m_1 - 0.01 \, \Msun} \right) 
    \end{split}
\end{equation}
where the last line applies for all our cases of interest, with $a > 0.5\Msun$.

For \textbf{BHBH}, we assume both primary and secondary masses range from $a = c = 20\Msun$ to $b = d = 300\Msun$:
\begin{equation}
    f_{\text{primary}} \times f_{\text{secondary}} = 5.05 \times 10^{-4}
\end{equation}

Similarly, for  \textbf{NSNS}, we assume both the primary and secondary range from $a = c = 8\,\Msun$ to $b = d = 20\,\Msun$, leading to: 
\begin{equation}
        f_{\text{primary}} \times f_{\text{secondary}}  \approx 8.12 \times 10^{-4}
\end{equation}

While for  \textbf{BHNS}  $a = 20\,\Msun$, $b = 300\,\Msun$ while $c = 8\,\Msun$ and $d = 20\,\Msun$:

\begin{equation}
        f_{\text{primary}} \times f_{\text{secondary}}  \approx 4.21 \times 10^{-4}
\end{equation}

\subsection{Initial separation}
We assume the ZAMS separation is independent of $m_1$ and $m_2$, i.e., $p(a ~|~m_1, ~m_2) = p(a)$.
We adopt a flat-in-log distribution of initial separations between $0.01$ and $1000\,\mathrm{AU}$. 
Using Figures \ref{fig: SSE properties}, and \ref{fig:max R per Z} we estimate separations avoiding mergers to be roughly \edited{$\gtrsim 0.1 \mathrm{AU}$} and the maximum for interaction to be approximately $14\mathrm{AU}$.
This leads to:

\begin{equation}
f_{\mathrm{init , sep}} = \frac{\log(14) - \log(0.1)}{\log(1000) - \log(0.01)} \approx 0.43
\end{equation}

\subsection{Approximate probability of being disrupted by a supernova}\label{app: analytic fsn}
Lastly, we examine the survival probability of the primary star following a supernova natal kick. Population synthesis studies are well-suited to estimate this probability, with typical values for $f_{\mathrm{survive , SN1}}$ around $14\%$ \citep[e.g.,][\tocite{Lam et al. in prep}]{2019A&A...624A..66R}. For our fiducial simulation, we find that $13-16\%$ of the systems survive the first supernova, with this range reflecting different metallicities. 
To align with the optimistic estimate we aim to compute, we adopt the upper limit from \tocite{Lam et al. (in prep)} for $f_{\mathrm{survive, SN1}} \approx 0.23$.

\subsection{Average stellar mass}\label{app: average Msun per bin}

To compute the average stellar mass per binary, $\langle M_{SF} \rangle$, we sample $N_{\mathrm{sys}}$  masses $m_1$, from a Kroupa IMF (equation \ref{eq: kroupa}).  
We subsequently bin $m_1$ using the following bin edges: $[0.05, 0.08, 0.5, 1, 10, 300]$. For each bin, we assume binary fractions of $[0.1, 0.25, 0.5, 0.75, 1]$, which are chosen to approximately follow Figure 1 from \citet{2023ASPC..534..275O}.
To draw secondary masses, we use $m_2 = q \cdot m_1$, and we again sample mass ratios $q$ from a uniform probability distribution.
This leads to 

\begin{equation}
\langle M_{SF} \rangle \approx 0.51 \Msun.
\end{equation}

\section{Remnant masses and Stellar radii in COMPAS}\label{app: single stars}
\begin{figure}
    \centering
        \includegraphics[width=0.49\textwidth]{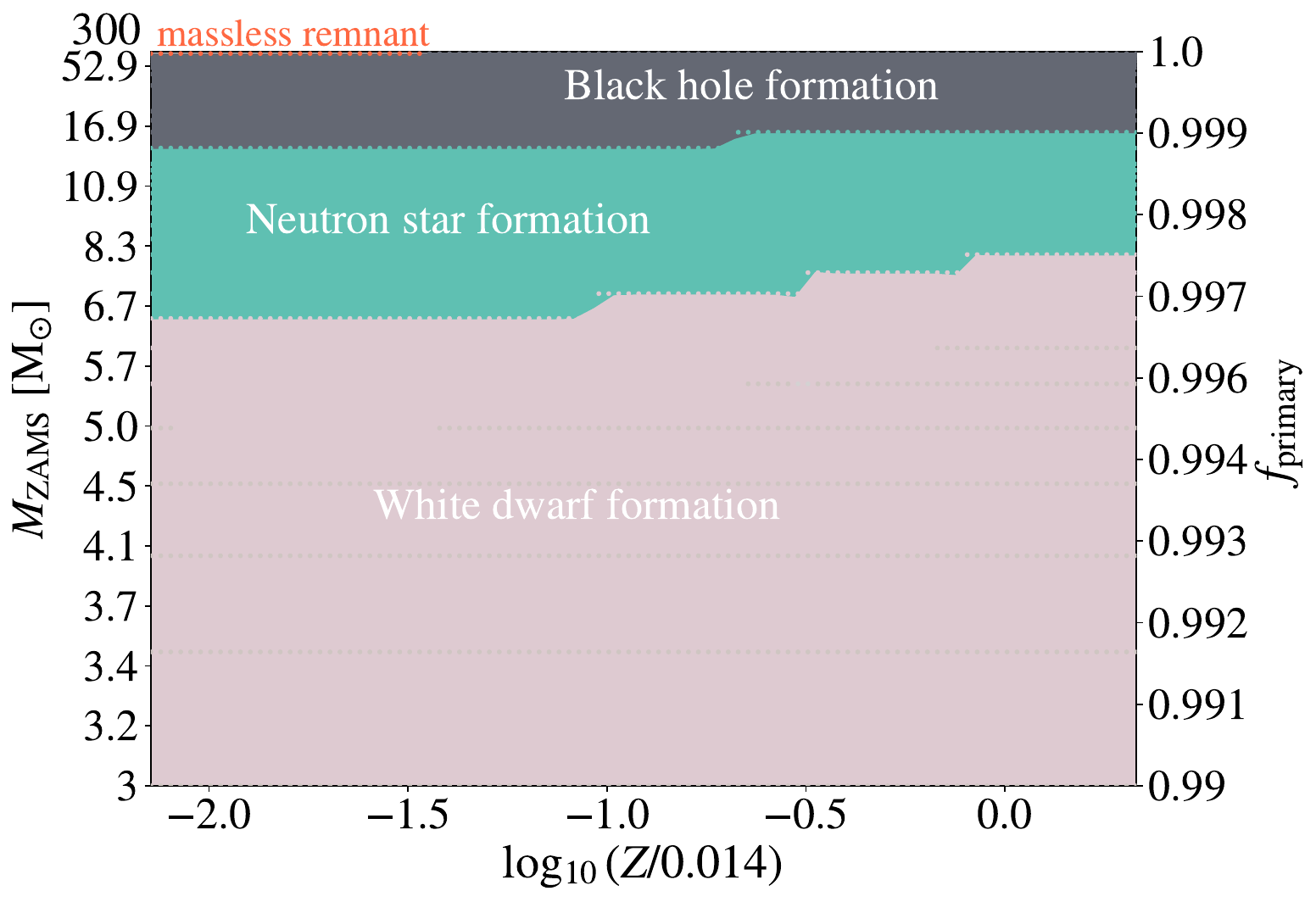}
    \caption{The approximate regions of WD, NS and BH formation from a ZAMS mass at different $Z$. The y-axis is scaled to follow a Salpeter IMF.
    \label{fig: rem masses} }
\end{figure}

In Figure \ref{fig: rem masses} we show the outcomes of single stars in COMPAS as a function of $M_{\mathrm{ZAMS}}$ and metallicity. We find that $f_{\mathrm{primary}} \approx 1-2 \times 10^{-3}$ for BH and NS formation. 

\begin{figure}
    \centering
    \includegraphics[width=0.55\textwidth]{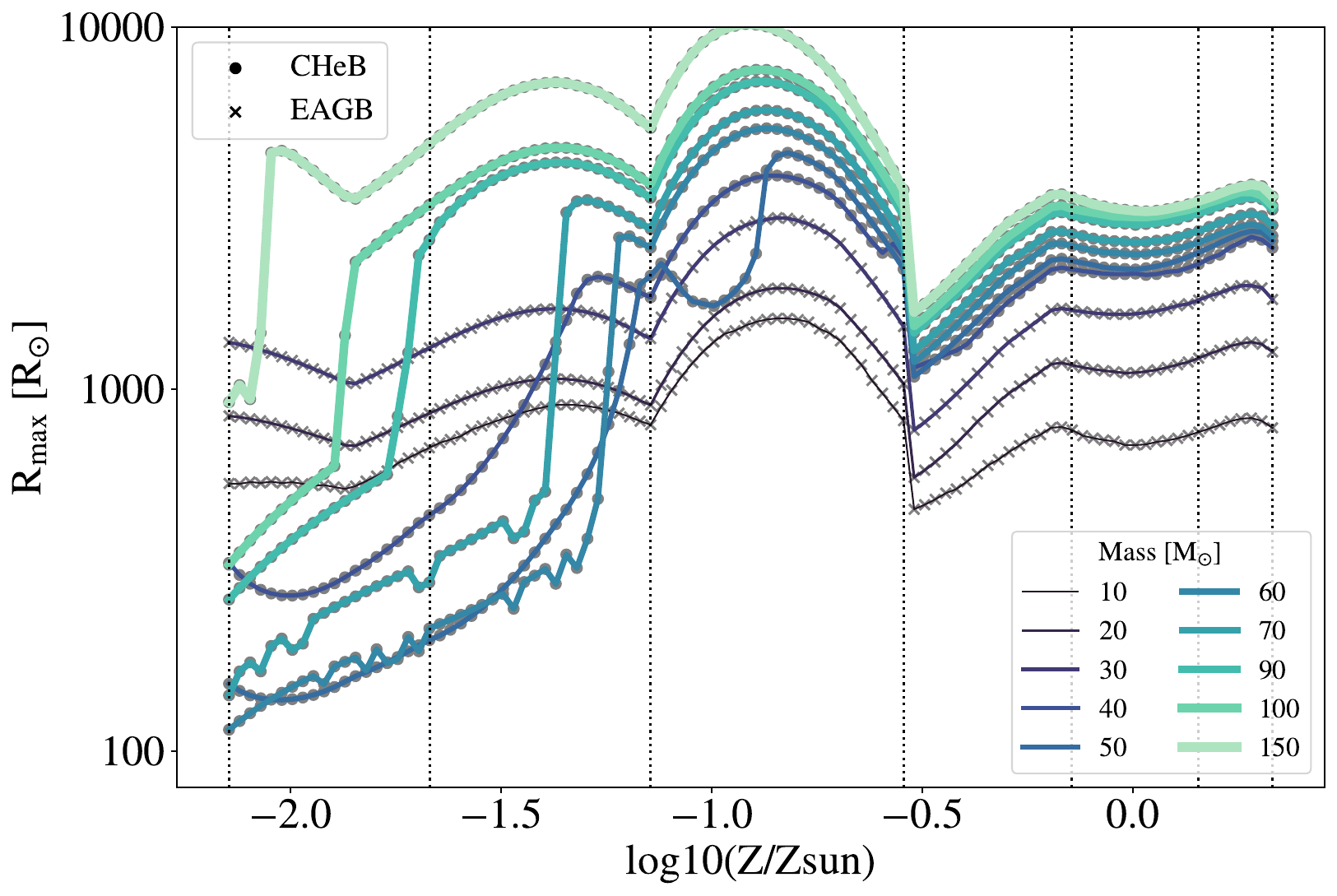}
    \caption{The maximum radius of stars at different metallicities. Markers indicate whether the star reaches $R_\mathrm{max}$ during core He burning (CHeB) or on the `Early AGB' phase (EAGB). Dotted lines show the metallicities of the original BSE code. }
    \label{fig:max R per Z}
\end{figure}

In Figure \edited{\ref{fig:max R per Z}}, we show the maximum radius of stars at different ZAMS masses as a function of metallicity. The plot exposes interpolation effects inherent to BSE-based codes, particularly a pronounced peak at approximately $\log_{10}(Z/\Zsun) \approx -0.9$ where stellar radii of massive stars extend up to $10^4\Rsun$. 
However, \cite{2023MNRAS.525..706R} argue that this does not significantly impact GW source properties, since most mass transfer events leading to GW sources are initiated before the donor star reaches its maximum radius. 
It is worth noting that at low metallicities, massive stars (with $M_\mathrm{ZAMS}>30\Msun$) remain more compact than lower-mass stars ($M_\mathrm{ZAMS} \leq 30\Msun$) due to our luminous blue variable mass-loss prescription.

\bibliography{main}
\bibliographystyle{aasjournal}


\end{document}